%% file: need.tex
\documentclass{LMCS}

\usepackage{enumerate}
\usepackage{hyperref}
\usepackage{natbib}
\usepackage{pdflscape} %
\let\cite=\citep

\input{definitions.tex}

\command{runCC[runCC\;],pushPrompt[pushPrompt\;],withSubCont[withSubCont\;],%
pushSubCont[pushSubCont\;],newPrompt[newPrompt],let[let\;],in[\;in\;],%
if[if\;],then[then\;],else[else\;]}

\newcommand{\mysf}[1]{\textsf{\small #1}}

\reservestyle{\term}{\mysf}

\term{letrec[letrec\;],be[\;be\;],lin[\;in\;]}

\newcommand{\topic}[1]{}

\newcommand{\consd}[2]{\ensuremath{\braket{#1,#2}}}
\newcommand{\flatten}[1]{\ensuremath{\mathcal{F}(#1)}}

\newcommand{\PP}[2]{\ensuremath{P^{#1}_{#2}}}
\newcommand{\LF}[0]{\<letrec>}
\renewcommand{\LF}[0]{\ensuremath{\mathsf{LR}\;}}

\def\doi{6 (3:1) 2010}
\lmcsheading%
{\doi}
{1--37}
{}
{}
{Oct.~12, 2009}
{Jul.~11, 2010}
{}

\begin{document}

\title{Lazy Evaluation and Delimited Control}

\author[R.~Garcia]{Ronald Garcia\rsuper a}
\address{{\lsuper a}Carnegie Mellon University}
\email{rxg@cs.cmu.edu}
\thanks{{\lsuper a}This work was supported by the National Science Foundation under
  Grant \#0937060 to the Computing Research Association for the CIFellows
  Project.}

\author[A.~Lumsdaine]{Andrew Lumsdaine\rsuper b} 
\address{{\lsuper{b,c}}Indiana University}
\email{\{lums,sabry\}@cs.indiana.edu}
\thanks{{\lsuper{b,c}}This work was supported by NSF awards CSR/EHS 0720857 and CCF 0702717}
\author[A.~Sabry]{Amr Sabry\rsuper c} 
\address{\vskip-6 pt}

\begin{abstract}

  The call-by-need lambda calculus provides an equational framework for
  reasoning syntactically about lazy evaluation.  This paper
  examines its operational characteristics.

  By a series of reasoning steps, we systematically unpack the standard-order
  reduction relation of the calculus and discover a novel abstract machine
  definition which, like the calculus, goes ``under lambdas.''  We prove that
  machine evaluation is equivalent to standard-order evaluation.

  Unlike traditional abstract machines, delimited control plays a significant
  role in the machine's behavior.  In particular, the machine replaces the
  manipulation of a heap using store-based effects with disciplined management
  of the evaluation stack using control-based effects.  In short, state is
  replaced with control.

  To further articulate this observation, we present a simulation of
  call-by-need in a call-by-value language using delimited control operations.
\end{abstract}

\maketitle

\section{Introduction}
\label{sec:intro}

\topic{Laziness and Control}
From early on, the connections between lazy
evaluation~\cite{friedman76cons,henderson76lazy} and control operations seemed
strong. One of these seminal papers on lazy evaluation~\cite{henderson76lazy}
advocates laziness for its coroutine-like behavior. Specifically, it motivates
lazy evaluation with a solution to the \emph{same fringe} problem: how to
determine if two trees share the same fringe without first flattening each tree
and then comparing the resulting lists.  A successful solution to the problem
traverses just enough of the two trees to tell that they do not match.  The
same fringe problem is also addressed in Sussman and Steele's original
exposition of the Scheme programming language~\cite{sussman98scheme}. One of
their solutions uses a continuation passing-style representation of coroutines.
More recently, \citet{Biernacki20057} explores a number of continuation-based
solutions to same fringe variants.

Same fringe is not the only programming problem that can be solved using either
lazy evaluation or continuations.  For instance, lazy streams and continuations
are also used to implement and reason about
backtracking~\cite{wand04backtracking,kiselyov05backtracking}.  Strong
parallels in the literature have long suggested that lazy evaluation elegantly
embodies a stylized use of coroutines. Indeed, we formalize this connection.

\topic{Reasoning about Call-by-need}
Call-by-need evaluation combines the equational reasoning capabilities of
call-by-name with a more efficient implementation technology that
systematically shares the results of some computations.  However,
call-by-need's evaluation strategy makes it difficult to reason about the
operational behavior and space usage of programs.  In particular, call-by-need
evaluation obscures the control flow of evaluation.  To facilitate reasoning,
semantic
models~\cite{launchbury93natural,sestoft97machine,friedman07krivine,nakata10need},
simulations~\cite{OKASAKI94}, and tracing tools~\cite{gibbons96tracing} for
call-by-need evaluation have been developed.
Many of these artifacts use an explicit store or store-based
side-effects~\cite{wang90lazy} to represent values that are shared between
parts of a program.  Stores, being amorphous structures, make it difficult to
establish program properties or analyze program execution.  This representation
of program execution loses information about the control structure of
evaluation. 
 
The call-by-need lambda calculus was introduced by~\citet{ariola95need} as an
alternative to store-based formalizations of lazy evaluation.  It is an
equational framework for reasoning about call-by-need programs and languages.
Following~\citet{plotkin75byname}, these authors present a calculus and prove a
standardization theorem that links the calculus to a complete and deterministic
(i.e. standard order) reduction strategy.  The calculus can be used to formally
justify transformations, particularly compiler optimizations, because any terms
it proves equal are also contextually equivalent under call-by-need evaluation.

Call-by-need calculi were investigated by two
groups~\cite{maraist98need,ariola97need}.  The resulting two calculi are quite
similar but their subtle differences yield trade-offs that are discussed in the
respective papers.  Nonetheless, both papers connect their calculi to similar
standard-order reduction relations.

One notable feature of Ariola and Felleisen's calculus (and both standard-order
reduction relations) is the use of evaluation contexts within the notions of
reduction.  It has been observed that evaluation contexts correspond to
continuations in some presentations of language
semantics~\cite{felleisen86secd,biernacka07framework}.  However, in these
systems evaluation contexts are used to model variable references and
demand-driven evaluation, not first-class continuations.

This paper exposes how Ariola et al.'s call-by-need evaluation relates
to continuations.  By systematically unpacking the standard-order reduction
relation of the calculus, we discover a novel abstract machine that models
call-by-need style laziness and sharing without using a store.  Instead, the
machine manipulates its evaluation context in a manner that corresponds to a
stylized use of delimited control operations.  The machine's behavior reveals a
connection between control operations and laziness that was present but hidden
in the reduction semantics.

To directly interpret this connection in the terminology of delimited control,
we construct a simulation of call-by-need terms in the call-by-value language
of~\citet{dybvig07monadic}, which provides a general framework for delimited
continuations with first-class generative prompts. 

Our concrete specifications of the relationship between call-by-need and
delimited control firmly establish how lazy evaluation relates to continuations
and other control-oriented language constructs and effects.
Implementations of both the machine and the simulation are available at the
following url: \\
\verb|http://osl.iu.edu/~garcia/call-by-need.tgz|.

\section{The Call-by-need Lambda Calculus}

The remainder of this paper examines Ariola and Felleisen's formalization of
call-by-need~\cite{ariola97need}.  The terms of the calculus are standard:
\begin{displaymath}
\begin{boxedarray}{lcl}
    t & \produce & x | \lambda x.t | t\;t \\
\end{boxedarray}  
\end{displaymath}
The call-by-need calculus, in direct correspondence with the call-by-value and
call-by-name lambda calculi, distinguishes lambda abstractions as values:
\begin{displaymath}
\begin{boxedarray}{lcl}
  v & \produce & \lambda x.t\\
\end{boxedarray}  
\end{displaymath}
Call-by-need is characterized by two fundamental properties: a computation is
only performed when its value is needed, and the result of any computation is
remembered and shared so that it only needs to be computed once.  This calculus
distinguishes two additional subsets of the term language to help represent
these properties.

To capture the notion of \emph{necessary} computations, the calculus
distinguishes the set of lambda abstractions that immediately need the value
of their argument.  To define this set, the calculus appeals to a notion of
evaluation contexts, a set of terms that each have a single hole ($[]$) in
them:
\begin{displaymath}
\begin{boxedarray}{lcl}
 E & \produce & [] | E\; t | (\lambda x.E[x])\;E | (\lambda x.E)\; t 
\end{boxedarray}  
\end{displaymath}
Given the evaluation contexts, the set of lambda abstractions in question is
defined syntactically as $(\lambda x.E[x])$, the set of lambda
abstractions whose bodies can be decomposed into an evaluation context and a
free instance of the abstracted variable.

The intuition for this definition is as follows.  Evaluation contexts are
essentially terms with a single \emph{hole} in them.  When used in the style
of~\citet{felleisen92revised}, evaluation contexts indicate those locations in
a program that are subject to evaluation. As such, a lambda abstractions of the
form $(\lambda x.E[x])$ will immediately refer to the value of its argument
when its body is evaluated.  Here, $E[x]$ is not a language construct, but
rather metalinguistic notation for a side condition on the term in the body of
the lambda abstraction.

The syntactic structure of call-by-need evaluation contexts give some hint to
the nature of computation captured by this calculus.  First, the context
production $E\;t$ indicates that evaluation can focus on the operator position
of an application regardless of the structure of the operand.  This property is
also true of call-by-name and call-by-value and reflected in their respective
evaluation contexts.  Second, the $(\lambda x.E[x])\;E$ production indicates
the operand of an application expression can be evaluated only if the operator
is a lambda abstraction that immediately needs its argument.  This restriction
does not hold for call-by-value\footnote{assuming left-to-right evaluation of
  application expressions}, where any lambda abstraction in operator position
justifies evaluating the operand.  This restriction on evaluation corresponds
with our intuitive understanding of call-by-need.  Third, the $(\lambda x.E)\;
t$ production indicates that evaluation can proceed \emph{under a lambda
  abstraction} when it is in operator position.  Though not immediately
obvious, this trait is used by the calculus to capture the process of sharing
computations among subterms.

To capture the notion of \emph{shared} computations, the call-by-need calculus 
distinguishes lambda abstractions with explicit bindings for some variables,
calling them \emph{answers}:
\begin{displaymath}
\begin{boxedarray}{lcl}
  a & \produce & v | (\lambda x.a)\;t
\end{boxedarray}  
\end{displaymath}
Answers are a syntactic representation of (partial) closures.  An answer takes
the form of a lambda term nested inside some applications.  The surrounding
applications simulate environment bindings for free variables in the nested
lambda term.
This representation makes it possible for the calculus to explicitly account
for variable binding and to syntactically model how call-by-need
evaluation shares lazily computed values.

The calculus has three notions of reduction:
\begin{displaymath}
\begin{boxedarray}{lcl@{}}
(\lambda x.E[x])\;v  & \notion &  (\lambda x.E[v])\;v \\
(\lambda x.a)\; t_1\; t_2  & \notion &  (\lambda x.a\; t_2)\; t_1 \\
(\lambda x_1.E[x_1])\;((\lambda x_2.a)\; t_1)  & \notion & 
        (\lambda x_2.(\lambda x_1.E[x_1])\;a)\; t_1 \\
\end{boxedarray}  
\end{displaymath}
The first reduction rule substitutes a value for a single variable instance in
an abstraction. The rule retains the binding and abstraction so as to share its
computation with other variable references as needed. The second and third
reduction rules commute an application with an answer binding to expose
opportunities for reduction without duplicating not-yet-needed computations.
These two rules help to ensure that computations will be shared among
references to a common variable.

As popularized by~\citet{barendregt}, each reduction assumes a hygiene
convention.  When combined with the evaluation contexts, the notions of
reduction yield a deterministic standard order reduction relation ($\sr$) and
its reflexive-transitive closure ($\Sr$).
\begin{defi}
  \label{def:sr}
  $t_1 \sr\, t_2$ if and only if $t_1 \equiv E[t_r]$, $t_2 \equiv E[t_c]$ 
  and $t_r \notion\ t_c$.
\end{defi}
Terms of the calculus satisfy unique decomposition, meaning that any
program (i.e. closed term) that is not an answer can be decomposed exactly one
way into a context $E$ and redex $t_r$. This property ensures that
$\sr$ is deterministic.
Standard order reduction is an effective specification of call-by-need
evaluation: if $t$ is a program (i.e. closed term), then $t$ call-by-need
evaluates to an answer if and only if $t \Sr \,a$ for some answer $a$.

\section{From Reduction Semantics to Machine Semantics}
\label{sec:reduction}

\topic{The point: Reduction Semantics don't immediately point to a
  tail-recursive implementation}

Some reduction semantics have been shown to correspond directly to abstract
machine semantics, thereby establishing the equivalence of a reducer and a
tail-recursive abstract machine
implementation~\cite{felleisen86secd,findler09redex}.  In particular,
\citet{danvyTRrefocusing} introduces a method and outlines criteria for
mechanically transforming reduction semantics into abstract machine semantics.
However, proceeding directly from the reduction semantics for call-by-need to a
tail-recursive abstract machine semantics poses some challenges that do not
arise with other reduction semantics like call-by-name and call-by-value.

A straightforward call-by-need reducer implementation na\"ively decomposes a
term into a context and a redex.  Any application could be one of three
different redexes, each of which is nontrivial to detect, so whenever the
decompose function detects an application, it sequentially applies several
recursive predicates to the term in hopes of detecting a redex.  If the term is
a redex, it returns; if not, it recursively decomposes the operator position.

Call-by-need redexes require more computational effort to
recognize than either call-by-name or call-by-value.  For instance, given a
term $t$, only a fixed number of terminal operations are required to detect
whether $t$ is a call-by-name redex: one to check if the term is an
application, one to access the operator position, and one to check if the
operator is an abstraction.

Contrast this with the call-by-need redex $(\lambda x.E[x])\:((\lambda
y.a)\:t)$.  Given a call-by-need term $t_x$, testing whether it matches this
redex form requires an unknown number of operations: check if~$t_x$ is an
application; check if its operator position is a lambda abstraction; check, in
an unknown number of steps, if the operator's body can be decomposed into
$E[x]$, where $x$ is both free in $E$ and bound by the operator; and finally
check, in an unknown number of steps, if the operand has the inductive
structure of an answer.

To make matters worse, some terms can be decomposed into the form $E[x]$ in more
than one way.  For instance, consider the term $(\lambda x.(\lambda y.y)\:x)$.
It can be decomposed as both $(\lambda x.E_1[y])$ and $(\lambda x.E_2[x])$
where $E_1\equiv (\lambda y.[])\:x$ and $E_2\equiv (\lambda y.y)\:[]$. As such,
a recursive procedure for decomposing a term cannot stop at the first variable
it finds: it must be able to backtrack in a way that guarantees it will find
the right decomposition $E[x]$---and in turn the right redex---if there is one.

Recall that one of the evaluation contexts has the form 
$(\lambda x.E)\:t$.  This means that redex evaluation can occur ``under
binders''~\cite{sabry04recursion,kameyama08closing}.
All three call-by-need notions of reduction shuffle lambda abstractions about
in unusual ways.  Furthermore, while reducing a recursive routine, a
call-by-need evaluator may end up performing reductions under multiple copies
of the same lambda abstraction.  Call-by-name and call-by-value evaluators can
address hygiene concerns by using environments and closures, but a call-by-need
evaluator must prevent its evaluation context from incorrectly capturing free
variable references.  Any evaluator that goes under lambdas must pay
particular attention to hygiene~\cite{xi97underlambda}.

Since this work was originally published, \citet{danvy10need} have adapted and
extended the method of \citet{danvyTRrefocusing} to produce a related
abstract machine for call-by-need.

\subsection{Towards an Abstract Machine}

To find call-by-need redexes tail-recursively, we apply an insight from
the CK abstract machine~\cite{felleisen86secd,findler09redex}.
The CK machine implements an evaluation strategy for call-by-value based on a
reduction semantics using the (inside-out) evaluation contexts $[], E[[]\;t]$
and $E[(\lambda x.t)\:[]]$.  To find a redex, the machine iteratively
examines the outermost constructor of a term and uses the evaluation context to
remember what has been discovered.  Since call-by-value satisfies unique
decomposition, this approach will find the standard redex if there is one.

To illustrate this in action, we walk through an example.  Consider the program
$(\lambda x.x)\:\lambda y.y$.  Evaluation begins with configuration
$\braket{[\;], \underline{(\lambda x.x)\:\lambda y.y}}$.  Underlining
indicates subterms that the machine knows nothing about;  at the beginning of
evaluation, it knows nothing about the entire term.  On
the first step of the reduction, the machine detects that the term is an
application $\underline{(\lambda x.x)}\;\underline{\lambda y.y}$. 
To examine the term further, the machine must move its focus to either the
operator or operand of this application. Since the machine is tail recursive,
it must also push an evaluation context to store the as-yet uncovered structure
of the term.  The only context it can reliably push at this point is
$[[]\;\underline{\lambda y.y}]$: it cannot push
$[\underline{(\lambda x.x)}\;[]]$ because it has not yet discovered that the
operator position is a value.  So the machine pushes the
$[[]\;\underline{\lambda y.y}]$ context, which serves as a reminder that it is
focused on the operator of an application.

On the second step of the reduction, the machine detects that the operator is
an abstraction $\lambda x.\underline{x}$, and observes that the innermost
context is $[[]\;\underline{\lambda y.y}]$.  In response, the machine pops the
context, focuses on $\underline{\lambda y.y}$, and pushes the context
$[(\lambda x.\underline{x})\;[]]$, since the operator is now known to be a
value.  This context serves as a reminder that evaluation is currently
focused on the operand of an application that can be reduced once that operand
becomes a value.

On the third step, the machine detects the abstraction 
$(\lambda y.\underline{y})$, and remembers that the innermost context is
$[(\lambda x.\underline{x})\;[]]$.  At this point, the machine has deduced
enough information to recognize the redex
$(\lambda x.\underline{x})\:\lambda y.\underline{y}$.
This example illustrates how the CK machine uses a depth-first left-to-right
search strategy to detect call-by-value redexes.

Now consider the same term under call-by-need using a similar strategy. As with
call-by-value, the top-level application can be detected, the operand can be
pushed onto the evaluation context, and the operator can be exposed as the
abstraction $\lambda x.\underline{x}$.  At this point behavior must diverge
from call-by-value because the body of the abstraction is still unknown and
call-by-need does not have
$[(\lambda x.t)\:[]]$ contexts for arbitrary $t$.  However, call-by-need does
have contexts of the form
$[(\lambda x.[])\:\underline{\lambda y.y}]$.
Therefore, it is possible to proceed under the first lambda abstraction, push
the context, and focus on $\underline{x}$.

The term is exposed as a variable $x$, which combines with the
context $[(\lambda x.[])\:\underline{\lambda y.y}]$ to form the term 
$(\lambda x.E[x])\;\underline{\lambda y.y}$ (where~$E\equiv[]$).  At this
point, enough information has been uncovered to push the context $[(\lambda
x.[][x])\;[]]$ and focus on $\underline{\lambda y.y}$.  The abstraction $\lambda
y.\underline{y}$ is recognized, and with that a call-by-need redex
$(\lambda x.[][x])\:\lambda y.\underline{y}$ has been found.
Success with this example suggests a promising strategy for implementing
call-by-need reduction tail-recursively.  

\subsection{An Initial Abstract Machine}
\label{sec:informal}
In this section, we elaborate the above search strategy into a simple but
inefficient tail-recursive abstract machine.  We present it without proof and
then by a series of correct transformations we derive an efficient machine that
we prove correct.

This abstract machine uses the same terms, values, and answers as the
calculus. However, it introduces two alternate notions.  First, the machine
uses a more versatile representation of evaluation contexts.  As observed
in~\citet{danvyTRrefocusing}, evaluation contexts can be mathematically
specified in more than one way. For optimal flexibility, we define evaluation
contexts as lists of frames, where the empty list $[\;]$ and single-frame lists
$[f]$ are our simple units, and the operator $\comp$ stands for list
concatenation.
\begin{displaymath}
\begin{boxedarray}{lcl}
f & ::= & []\; t | (\kappa x.E)\;[] | (\lambda x.[])\; t \\[0.5ex]
E & ::= & [\;] | [f] \comp E | E \comp [f] \\[0.5ex]
& &\text{where } E \comp [\;] = [\;] \comp E = E\\[0.5ex]
& &\text{and } E_1 \comp (E_2\comp E_3) = (E_1\comp E_2)\comp E_3\\
\end{boxedarray}  
\end{displaymath}
When two contexts are composed, the second context is plugged into the hole of
the first context: for example
$[[]\; t_2] \comp [[]\; t_1] = [([]\; t_1)\;t_2]$.

We call the frame $[(\lambda x.[])\: t]$ a \emph{binder} frame. It represents a
variable binding in the context.  It can be read as 
$[\mathrm{let}\,x=t\;\mathrm{in}\;[]]$, but we use the former notation to
emphasize that call-by-need evaluation proceeds under lambdas.  This
observation motivates our analysis of hygiene in Section~\ref{sec:hygiene}.

We call the frame $[(\kappa x.E)\;[]]$ a \emph{cont} frame, in reference to
continuations.  The construction $(\kappa x.E)$ is called a cont and replaces
the metalinguistic term notation $(\lambda x.E[x])$ from the calculus.  We use
a different notation for conts than lambda terms to indicate that in the
machine conts are distinct from terms (they are of type \texttt{Cont} rather
than type \texttt{Term} in an implementation).  Conts indicate nontrivial
structural knowledge that the machine retains as it searches for a redex.  This
distinction matters when we establish continuation semantics for
machine states.  As we shall see, a cont frame represents a suspended variable
reference.

Finally we call the frame $[[]\;t]$ an \emph{operand} frame, and it represents
a term waiting for an abstraction.

The abstract machine also introduces a notion of \emph{redexes}:
\begin{displaymath}
\begin{boxedarray}{lcl}
 r & \produce & a\;t | (\kappa x.E)\;a 
\end{boxedarray}  
\end{displaymath}
Redexes are distinguished from terms in the machine, meaning that in an
implementation, the type \texttt{Redex} is distinct from the type \texttt{Term}.
This distinction suggests that conts $(\kappa x.E)$
are neither terms nor first-class entities in the call-by-need language: they
only appear in evaluation contexts and in redexes.  As we discuss below, the
machine distinguishes one more kind of redex than the calculus.

The transition rules for the machine are staged into four distinct
groups: refocus, rebuild, need, and reduce.  Each machine configuration can
be related to a term in the language of the calculus.
The \emph{refocus} rules examine the current term and push as many
operand frames $[[]\;t]$ as possible.  A refocus configuration $\refocus{E,t}$
represents the term $E[t]$.
\begin{displaymath}
  \begin{boxedarray}{@{}l@{}}
     \vspace{1ex}
    \boxed{\refocus{E,t}}\quad\desc{Refocus} \\
    \begin{array}{rcl}
      \refocus{E, x}
      & \mapsto &
      \need{E,[\;],x} \\[0.5ex]

      \refocus{E, \lambda x.t}
      & \mapsto &
      \rebuild{E, \lambda x.t} \\[0.5ex]

      \refocus{E, t_1\;t_2}
      & \mapsto &
      \refocus{E \comp [[]\;t_2], t_1} 
    \end{array}
\end{boxedarray}  
\end{displaymath}
Upon reaching a variable, refocus transitions to the need rules; upon reaching a
lambda abstraction, it transitions to the rebuild rules. 

The \emph{rebuild} rules search up into the context surrounding
an answer for the next applicable redex. A rebuild configuration
$\rebuild{E,a}$ represents the term $E[a]$.
\begin{displaymath}
  \begin{boxedarray}{@{}l@{}}
     \vspace{1ex}
    \boxed{\rebuild{E,a}}\quad\desc{Rebuild} \\
    \begin{array}{rcl}
      \rebuild{[\;],a} & \mapsto & a \\[0.5ex]

      \rebuild{E \comp [[]\; t_1], a} & \mapsto &
      \reduce{E, a\; t_1} \\[0.5ex]
      
      \rebuild{E \comp [(\lambda x.[])\; t_1], a} & \mapsto &
      \rebuild{E, (\lambda x.a)\; t_1} \\[0.5ex]

      \rebuild{E_1 \comp [(\kappa x.E_2)\; []], a} & \mapsto &
      \reduce{E_1, (\kappa x.E_2)\; a}
    \end{array}
  \end{boxedarray}
\end{displaymath}  
These rules examine the current context and proceed to build a maximal
answer-shaped term, progressively wrapping binder frames around the current
answer.  If the entire context is consumed then evaluation has completed and
the entire program is an answer.  Upon reaching an operand or cont frame, a
redex has been found, and rebuild transitions to the reduce rules.  These rules
resemble the $\mathrm{refocus}_{\mathrm{aux}}$ rules of~\citet{danvyTRrefocusing}.

The \emph{need} rules also examine the context, but they
search for the binder frame that corresponds to the variable under focus.
A need configuration $\need{E_1,E_2,x}$ represents the term $E_1[E_2[x]]$.
\begin{displaymath}
  \begin{boxedarray}{@{}l@{}}
     \vspace{1ex}
    \boxed{\need{E,E,x}}\quad\desc{Need} \\
    \begin{array}{rcl}
      \need{E_1 \comp [(\lambda x.[])\; t], E_2, x} & \mapsto &
      \refocus{E_1 \comp [(\kappa x.E_2)\; []], t} \\[0.5ex]
      \need{E_1 \comp [f], E_2, x} & \mapsto &
      \need{E_1,[f] \comp E_2, x} \\[0.5ex]
      \multicolumn{3}{c}{\text{where, }[f] \not\equiv [(\lambda x.[])\; t]}
    \end{array}
  \end{boxedarray}
\end{displaymath}  

Since input programs are closed, the associated binder must be somewhere in the
context.  Upon finding the right binder frame, a cont frame $[(\kappa x.E)\;
[]]$ is pushed onto the context and evaluation proceeds to refocus on the
operand from the associated binder frame.

The \emph{reduce} rules simulate the notions of reduction from the calculus.
A reduce configuration $\reduce{E,r}$ represents the term $E[r]$ where a cont
$\kappa x.E$ represents the term $\lambda x.E[x]$.
\begin{displaymath}
  \begin{boxedarray}{@{}l@{}}
     \vspace{1ex}
    \boxed{\reduce{E,r}}\quad\desc{Reduce} \\[0.5ex]
    \begin{array}{rcl}
      \reduce{E_1, (\kappa x.E_2)\; v}
      & \mapsto &
      \refocus{E_1, (\lambda x.E_2[v])\; v} \\[0.5ex]

  \reduce{E_1, (\kappa x_1.E_2)\;((\lambda x_2. a)\; t)}& \mapsto
  & \refocus{E_1, (\lambda x_2.(\lambda x_1.E_2[x_1])\; a)\; t} \\[0.5ex]

      \reduce{E, (\lambda x. a)\; t_1\; t_2}
      & \mapsto &
      \refocus{E, (\lambda x. a\;t_2)\; t_1} \\[0.5ex]

      \reduce{E, (\lambda x.t_1)\;t_2}
      & \mapsto &
      \refocus{E \comp [(\lambda x.[])\; t_2], t_1}  \\[0.5ex]
    \end{array} 
\end{boxedarray}
\end{displaymath}

Each of the first two reduce rules transforms a cont into a lambda abstraction
by plugging its context with a term and abstracting its variable.  As such,
each reduce rule transforms a redex into a pure term of the calculus and
transitions to a refocus configuration, which searches for the next redex.

The reduce rules also handle terms of the form $(\lambda x.t_1)\;t_2$, even
though such terms are not call-by-need redexes.  Including this rule gives the
set of redexes greater uniformity: all terms of the form $a\;t$ are redexes,
just like the terms of the form $(\kappa x.E)\; a$.  This symmetry is not
exhibited in the call-by-need calculus.  However, \citet{ariola97need} defines
and uses an auxiliary \emph{let calculus} that adds the reduction
\begin{displaymath}
      (\lambda x.t_1)\;t_2 \notion\ \mathrm{let}\; x=t_2\; \mathrm{in}\; t_1
\end{displaymath}
to the calculus and defines the other reductions in terms of the $\mathrm{let}$
expressions.  The fourth reduce rule corresponds to this reduction rule.
However, our presentation shows that an auxiliary $\mathrm{let}$ term, though
compatible with this model, is not needed to specify call-by-need: the syntax
of pure calculus terms suffices.  Furthermore, treating this rule as a
reduction here anticipates a change we make to the machine
(Section~\ref{sec:hygiene}) that adds explicit variable renaming to that rule.
Finally, the reduce rules are improved in the next section so that all reduce
rules change their representative terms nontrivially.

Machine evaluation of a program $t$ begins with the refocus configuration
$\refocus{[\;],t}$ and terminates if it arrives at an answer $a$.
Its behavior in between can be summarized as follows: search downwards until a
value or variable reference is reached.  If a variable reference is reached,
store a cont in the context to remember the variable reference and proceed to
evaluate its binding.  If an abstraction is reached, accumulate an answer up to
the innermost redex, or the top of the evaluation context if none is found.  In
short, the machine performs a depth-first, left-to-right traversal in search of
a call-by-need redex.  Along the way it uses the evaluation context to store
and retrieve information about program structure, particularly the location of
variable bindings (using binder frames) and variable references (using cont
frames). The refocus, rebuild, and need rules leave the term representation of
their configurations unchanged (e.g. if $\refocus{E_1,t_1} \mapsto
\refocus{E_2,t_2}$ then $E_1[t_1] \equiv E_2[t_2]$), and the reduce rules
embody the notions of reduction from the calculus.

Our strategy for producing this machine builds on the strategy of
\citet{danvyTRrefocusing}, which mechanizes the direct transformation of
reduction semantics into abstract machine semantics.  That report introduces
and verifies a general method for using reduction semantics that meet certain
criteria to construct a function that ``refocuses'' an arbitrary term-context
pair to a redex-context pair. The resulting function can then be used to induce
an abstract machine semantics.  Unfortunately that refocus function
construction does not apply to the call-by-need lambda calculus because
the calculus does not meet the required criteria.  In particular, the
construction requires that a maximally-decomposed closed term (i.e. program)
will focus on a value or a redex.  However, call-by-need evaluation contexts
can decompose down to variable references, which are neither redexes nor values
under call-by-need.
There are however other ways to produce an abstract machine from a reduction
semantics which may apply to the call-by-need calculus studied here.  For
instance, \citet{danvy10need} devise a variant of the let-based call-by-need
reduction semantics, implement it, and use a program-transformation based
approach to produce a refocus function and abstract machine implementation.

The following partial trace demonstrates how the initial abstract machine
discovers the first redex for our running example $(\lambda x.x)\;\lambda y.y$:
\begin{displaymath}
\begin{array}{llllll}
& \braket{[\;], \underline{(\lambda x.x)\;\lambda y.y}}_f & \mapsto &
\refocus{[[]\;\underline{\lambda y.y}], \underline{\lambda x.x}} & \mapsto &
\rebuild{[[]\;\underline{\lambda y.y}], \lambda x.\underline{x}} \\[1.0ex]
 \mapsto &
\reduce{[\;], (\lambda x.\underline{x})\;\underline{\lambda y.y}} & \mapsto &
\refocus{[(\lambda x.[])\;\underline{\lambda y.y}], \underline{x}} & \mapsto &
\need{[(\lambda x.[])\;\underline{\lambda y.y}],[\;], x} \\[1.0ex]
 \mapsto & 
\refocus{[(\kappa x.[\;])\;[]], \underline{\lambda y.y}} & \mapsto &
\rebuild{[(\kappa x.[\;])\;[]], \lambda y.\underline{y}} & \mapsto & 
\reduce{[\;], (\kappa x.[\;])\;\lambda y.\underline{y}} 
\end{array}
\end{displaymath}

\section{Refining the Machine}

In this section we study the behavior of the abstract machine and
make some improvements based on our observations.  These changes lead us
from the initial machine above to our final machine specification.

\subsection{Grabbing and Pushing Conts}
\label{sec:conts}
The need rules linearly search the nearest context for a binder
frame that matches the variable under question.  This process can be specified
as one step:
\begin{gather*}
  \need{E_1 \comp [(\lambda x.[])\; t] \comp E_2, x}  \mapsto
\refocus{E_1 \comp [(\kappa x.E_2)\; []], t} \\
\text{where } \lbrack(\lambda x.[])\; t] \notin E_2
\end{gather*}
This evaluation step accumulates a segment of the current evaluation context
and stores it.  In general, abstract machines that model control operators
represent control capture in a similar manner.  In this particular case, only
part of the evaluation context is captured, and the amount of context captured
depends on the dynamic location in the context of a certain frame.  As such,
the need rules seem to perform some kind of \emph{delimited} control
capture. This analogy becomes stronger upon further analysis of the first
reduce rule from Section~\ref{sec:informal}.
The machine uses its structural knowledge of $\kappa x.E$ to construct the
abstraction $\lambda x.E[v]$.  However, the resulting machine configuration no
longer retains any of the structure that had previously been discovered.
Recall our example execution trace from Section~\ref{sec:informal}. 
The machine reduces the redex found at the end of that trace as follows:
\begin{gather*}
  \reduce{[\;], (\kappa x.[\;])\;\lambda y.\underline{y}} \mapsto
  \braket{[\;],\underline{(\lambda x.\lambda y.y)\;\lambda
      y.y}}_f
\end{gather*}
By returning to refocus following the reduction, the machine loses all
structural knowledge of the term.  To continue execution, it must
examine the structure of the contractum from scratch.
Fortunately, the evaluator can be safely improved so that it retains 
knowledge of the contractum's structure:
\begin{prop}
  \begin{equation*}
     \refocus{E_1, (\lambda x.E_2[v])\;v}  \mapsTo 
     \rebuild{E_1 \comp [(\lambda x.[])\;v] \comp E_2, v} 
  \end{equation*}
\end{prop}
\begin{proof}
  Corollary of $\refocus{E_1, E_2[v]} \mapsTo \rebuild{E_1\comp E_2,v}$,
  which is proven by induction on $E_2$. 
\end{proof}
This proposition justifies replacing the first reduce rule with one that pushes the evaluation context embedded
in the cont and proceeds to rebuild an answer:
\begin{gather*}
  \reduce{E_1, (\kappa x.E_2)\; v}
  \mapsto 
  \rebuild{E_1 \comp [(\lambda x.[])\; v] \comp E_2, v} 
\end{gather*}
This short-circuit rule extends the current evaluation context with a binder
frame and the context $E_2$ that was inside the cont.  The rule is suggestive
of delimited control because machine models of control operators generally
represent the reinstatement of delimited continuations by extending the current
context with a piece of captured evaluation context.  Of more immediate
interest, though, is how reduction of our example now proceeds:
\begin{gather*}
  \reduce{[\;], (\kappa x.[\;])\;\lambda y.\underline{y}} \mapsto
  \rebuild{[(\lambda x.[])\;(\lambda y.\underline{y})],\lambda y.\underline{y}} 
\end{gather*}
All knowledge of the contractum's structure is retained, though much of it is
now stored in the evaluation context.

\subsection{Shifting Binder Frames}

The second and third reduce rules from Section~\ref{sec:informal} also discard
structural information.  Specifically, they both transition to the forgetful
refocus rule.  However their information can be preserved.
\begin{prop}
  \begin{equation*}
      \refocus{E, (\lambda x.a\;t_2)\; t_1} \mapsTo
      \reduce{E \comp [(\lambda x.[])\;t_1], a\;t_2}.
  \end{equation*}
\end{prop}
\begin{proof}
  Corollary of $\refocus{E_1, a} \mapsTo \rebuild{E_1,a}$, which is proven by
  induction on $a$.
\end{proof}
\begin{prop}
  If $E_2$ does not capture $x_1$ (Section~\ref{sec:hygiene}), then
  \begin{equation*}
    \refocus{E_1,(\lambda x_2.(\lambda x_1.E_2[x_1])\; a)\; t} \mapsTo 
    \reduce{E_1 \comp [(\lambda x_2.[])\; t], (\kappa x_1.E_2)\; a}.
  \end{equation*}
\end{prop}
\begin{proof}
  Corollary of $\refocus{E_1, a} \mapsTo \rebuild{E_1,a}$ and \\
    $\refocus{E_1,(\lambda x_1.E_2[x_1])\; t} \mapsTo
    \refocus{E_1 \comp [(\kappa x_1.E_2)\; []], t} $, which is proven by case
    analysis and induction on $E_2$.
\end{proof}
These propositions justify short-circuiting the respective evaluation rules.
The new rules improve the behavior of the abstract machine.
\begin{equation*}
\reduce{E_1, (\kappa x_1.E_2)\;((\lambda x_2. a)\; t)} \mapsto
\reduce{E_1 \comp [(\lambda x_2.[])\; t], (\kappa x_1.E_2)\; a}
\end{equation*}
\begin{gather*}
\reduce{E, (\lambda x. a)\; t_1\; t_2}
\mapsto 
\reduce{E \comp [(\lambda x.[])\;t_1], a\;t_2}
\end{gather*}
By fast-forwarding
to reduce, the rules retain the discovered term structure and thereby
avoid retracing the same terms again.

\subsection{Answer \texorpdfstring{$=$}{=} Binders \texorpdfstring{$\times$}{x} Value}

The transition rules repeatedly create binder frames out of terms and reabsorb
those frames into answers.  In this section we simplify this protocol.
We distinguish answers from terms by providing them a separate
representation:
\begin{displaymath}
\begin{boxedarray}{lclr}
  a & \produce & \answer{E,v}, &\text{where }
  E = \overline{[(\lambda x_i.[])\;t_i]} \\
\end{boxedarray}  
\end{displaymath}
An answer is now represented as a tuple containing the nested lambda
abstraction and the sequence of binder frames that are wrapped around it in the
old presentation (we use overlines to indicate sequences).  This presentation
bears strong similarity to calculi with explicit
substitutions~\cite{abadi91explicit} in that each binder frame $[(\lambda
x.[])\;t]$ corresponds to a substitution $[t/x]$.  An answer can be seen as a
lambda term nested inside a sequence of explicit substitutions,
$v\overline{[t_i/x_i]}$.

The rebuild rules could be reformulated as a three place configuration,
$\rebuild{E,E,v}$, but instead we immediately apply the same improvement that
we applied to the need rules in Section~\ref{sec:conts}. For instance, the new
transition rule for rebuilding to a cont frame is:
\begin{gather*}
  \rebuild{E_1 \comp [(\kappa x.E_2)\; []] \comp E_3, v} \mapsto
\reduce{E_1, (\kappa x.E_2)\; \answer{E_3,v}} \\
\text{where } E_3 = \overline{[(\lambda x_i.[])\;t_i]}
\end{gather*}

Returning to our running example, reduction from its most recent state (at the
end of Section~\ref{sec:conts}) transitions to a final answer, signaling the
end of execution:
\begin{gather*}
  \rebuild{[(\lambda x.[])\;\lambda y.\underline{y}],\lambda y.\underline{y}}
  \mapsto
  \Braket{\answer{[(\lambda x.[])\;\lambda y.\underline{y}],
      \lambda y.\underline{y}}}
\end{gather*}

\subsection{Aggregate Reduction}

Now that answers explicitly carry their binders in aggregate, the reduce
rules can be substantially consolidated.  Currently, the second and third
reduce rules iteratively remove the top binder frame from an answer and
push it onto the evaluation context.  This process repeats until the answer
is just a lambda abstraction.  At that point, the second and third reduce
rules defer to the first and fourth reduce rules respectively. This
corresponds exactly with standard-order reduction 
(cf. Definition~\ref{def:sr}):
\begin{prop}
\label{prp:fast-sr}
\begin{equation*}
  \begin{split}
  & E[((\lambda x_n.\, \dots 
  ((\lambda x_1. ((\lambda x_0.v)\; t_0)) \;t_1) \;\dots)\;t_n) \; t] 
 \Sr \\
  & E[((\lambda x_n.\, \dots 
  ((\lambda x_1. ((\lambda x_0.v\; t)\; t_0)) \;t_1) \;\dots)\;t_n)].
  \end{split}
\end{equation*}

\begin{equation*}
  \begin{split}
    & E[(\lambda x.E[x])\;
    ((\lambda x_n.\, \dots 
    ((\lambda x_1. ((\lambda x_0.v)\; t_0)) \;t_1) \;\dots)\;t_n)]  \Sr  \\
    & E[((\lambda x_n.\, \dots ((\lambda x_1. ((\lambda x_0.
    (\lambda x.E[x])\; v)\; t_0)) \;t_1) \;\dots)\;t_n)].
  \end{split}
\end{equation*}
\end{prop}
\begin{proof}
  By induction on the structure of the answer term, using the
  unique decomposition lemma of~\citet{ariola97need}.
\end{proof}

Using the new answer representation, 
each pair of associated reduce rules can be merged into one omnibus rule
that moves all the binder frames at once and simultaneously performs a 
reduction using the underlying value. 
\begin{gather*}
  \reduce{E_1, (\kappa x.E_2)\; \answer{E_3, v}} \mapsto 
  \rebuild{E_1 \comp E_3 \comp [(\lambda x.[])\;v] \comp E_2, v} \\
  \reduce{E_1, \answer{E_2,(\lambda x.t_1)}\;t_2} \mapsto
  \refocus{E_1 \comp E_2 \comp [(\lambda x.[])\; t_2], t_1} 
\end{gather*}

As a result of these transformations, both conts and answers contain evaluation
contexts.  Furthermore, conts and answers are not terms of the calculus, and
the machine never reverts a cont or answer to a term.  The rules that create
them, rebuild for answers and need for conts, capture part of the evaluation
context, and the rules that consume them, the reduce rules, reinstate the
captured contexts.

\subsection{Variable Hygiene}
\label{sec:hygiene}

Presentations of calculi often invoke a hygiene convention and from then on pay
little attention to bound or free variables.  In this manner, calculi do not
commit to any of the numerous ways that hygiene can be enforced.  Many abstract
machines, however, use environments or explicit sources of fresh names to
guarantee hygiene and thereby provide a closer correspondence to concrete
implementations.  In this section, we augment the call-by-need machine with
simple measures to enforce hygiene.

Our primary hygiene concerns are that evaluation occurs under binders and
binders are shifted about in unusual ways.  In order to ensure that binding
structure is preserved throughout evaluation, we need to be able to reason
locally, within each machine configuration, about bound variables.  To make
this possible, we make one last change to the machine.  We add 
a list of names to each machine configuration.  
\begin{displaymath}
\begin{boxedarray}{lcl}
      X & \produce & \overline{x_i}\\
\end{boxedarray}  
\end{displaymath}
Most of the machine rules simply pass the list of names along to the next
transition. One of the reduce rules manipulates the list of names.
\begin{gather*}
      \tag{D.2} \reduce{X | E_1, \answer{E_2,\lambda x.t_1} \; t_2} \nam\
      \refocus{X,x' | E_1 \comp E_2 \comp [(\lambda x'.[])\;t_2], 
        t_1[x'/x]} \quad x' \notin X
\end{gather*}
When this rule goes under a lambda, it adds the name of its bound variable to
the list $X$ of variables.  The notation $X,x$ expresses adding a new name $x$
to $X$.  If the bound variable $x$ on the left hand side of the rule is already
a member of $X$, then the variable is renamed as part of the transition.  As
such, $X$ can be considered a set.

Now each machine configuration has one of five forms:
\begin{displaymath}
\begin{boxedarray}{lcl}
      \Braket{X|E,?} & \produce & \Braket{X|\answer{E,v}} | \reduce{X|E,r} | 
      \refocus{X|E,t} \\[0.5ex]
        & | & \rebuild{X|E,v} | \need{X|E,x}
\end{boxedarray}  
\end{displaymath}
We use the notation $\Braket{X|E,?}$ below to uniformly discuss all
configuration types, where $X$ refers to the list of names, $E$ refers to the
context, and $?$ refers to the term or redex.  For a final configuration
$\Braket{X|\answer{E,v}}$, $?$ refers to the answer's underlying value $v$, and
$E$ corresponds to the answer's binder frames $E$. We use the metavariable
$C$ to range over configurations when the internal structure does not matter.

\pagebreak
The call-by-need abstract machine uses the set $X$ of names to keep track of
\emph{active variables}: any variable $x$ whose binding instance has
been pushed into a binder frame $[(\lambda x.[])\; t]$:
\begin{displaymath}
\begin{boxedarray}{lcl} 
  AV([\;]) & = & \emptyset \\[0.5ex]
  AV([(\lambda x.[])\;t] \comp E) & = & \set{x} \union AV(E) \\[0.5ex]
  AV([[]\;t] \comp E) & = & AV(E) \\[0.5ex]
  AV([(\kappa x.E_1)\;[]] \comp E) & = &
  AV(E_1) \union \set{x} \union AV(E)
\end{boxedarray}
\end{displaymath}
Cont-bound variables are counted among the active variables because machine
evaluation must have gone under a binding to construct the cont frame.

The renaming condition on the $(D.2)$ reduce rule ensures that active
variables are mutually distinguishable.  This guarantees that the machine's
need rule can never capture the wrong evaluation context and thus execute the
wrong bound expression.

Renaming is not obviously sufficient to ensure bound variable hygiene because
of how the machine manipulates evaluation contexts.  For instance, even though
the need rule is guaranteed to only match a variable with the right binder
frame, we have no guarantee that the right binder frame could never be trapped
inside a cont frame and hidden from view while a need transition searches for
it.  Were this to happen, the machine would get stuck.  Furthermore, the
reduction rules flip and shift evaluation contexts that might contain binder
frames.  If a binder frame were to end up below another context frame that
contains references to its bound variable, then those references would no
longer be bound in the context; the need rule would exhaust the evaluation
context if it attempted to resolve any of these references.

To verify that machine evaluation is well-formed, we establish well-formedness
conditions that suffice to ensure safe evaluation and we show that they hold 
for evaluation of all programs.
The well-formedness conditions rely on straightforward notions
of captured context variables ($CV$) and 
free context variables ($FV$):
\begin{displaymath}
\begin{boxedarray}{lcl}
  CV([\;]) & = & \emptyset \\[0.5ex]
  CV(E \comp [(\lambda x.[])\;t]) & = & \set{x} \union CV(E)  \\[0.5ex]
  CV(E \comp [[]\;t]) & = & CV(E)  \\[0.5ex]
  CV(E \comp [(\kappa x.E_1)\;[]]) & = & CV(E) \\[0.5ex]
  \\
  FV([\;]) & = & \emptyset \\
  FV([(\lambda x.[])\;t] \comp E) & = & FV(t) \union (FV(E) - \set{x}) \\
  FV([[]\;t] \comp E) & = & FV(E) \union FV(t) \\
  FV([(\kappa x.E_1)\;[]] \comp E) & = & FV(E) \union (FV(E_1) - \set{x})
\end{boxedarray}
\end{displaymath}
As expected, these two notions are related.
\begin{lem}
\label{lm:fvcv}
\mbox{}
\begin{enumerate}[\em(1)]
\item $FV(E_1) \subseteq FV(E_1 \comp E_2)$.
\item $CV(E_1 \comp E_2) = CV(E_1) \union CV(E_2)$.
\item $FV(E_1 \comp E_2) = FV(E_1) \union (FV(E_2)\backslash CV(E_1))$.
\end{enumerate}
  
\end{lem}
\begin{proof}
  By induction on the length of $E_1$, $E_2$, and $E_1$ respectively.
\end{proof}

To establish that binder frames remain properly positioned, we define a notion
of well-formed evaluation contexts:

\begin{defi}
  A Machine context/name pair is well formed, notation $X | E \<wf>$, iff:
  \begin{enumerate}[(1)]
  \item $FV(E) = \emptyset$;
  \item $AV(E) = \set{x | x \text{ occurs in } X}$
  \item Each active variable of $E$ is distinct:
    if $E_1 \comp E_2 \in E$ then $AV(E_1) \cap AV(E_2) = \emptyset$, and
    if $[(\kappa x.E_1)\;[]] \in E$ then $x \notin AV(E_1)$.
  \end{enumerate}
\end{defi}
These well-formedness criteria ensure that a context has no unbound variable
references, that active variables cannot interfere with each other, and that $X$
is simply a particular ordering of the $E$'s active variables.  
The captured variables of $E$ are also distinct since every captured variable is
an active variable.

Furthermore, machine configurations also have a notion of well-formedness:
\begin{displaymath}
  \begin{boxedarray}{@{}l@{}}
    \begin{inferbox*}
      \inference{X|E\<wf> & FV(?) \subseteq CV(E)}
      {\Braket{X|E,?}_c \<wf>} \quad c \not\equiv d
      \and
      \inference{X|(E_1\comp [(\kappa x.E_2)\;[]] \comp E_3) \<wf> &
        FV(v) \subseteq CV(E_1 \comp [(\kappa x.E_2)\;[]] \comp E_3)}
      {\reduce{X|E_1,(\kappa x.E_2)\;\answer{E_3,v}} \<wf>}
      \and
      \inference{X|(E_1 \comp [[]\;t] \comp E_2) \<wf> &
        FV(v) \subseteq CV(E_1 \comp [[]\;t] \comp E_2)}
      {\reduce{X|E_1,\answer{E_2,v}\;t_2} \<wf>}
    \end{inferbox*}    
  \end{boxedarray}
\end{displaymath}

These well-formedness conditions ensure that the evaluation contexts $E$ are
well-formed, that the list of names $X$ matches the active variables of $E$,
and that the free variables of the term under focus are captured by the
context.  To account for redexes, the well-formedness conditions for each
reduce configuration reflect the well-formedness conditions for the
corresponding rebuild configuration.  As shown below, well-formed reduce
configurations $\reduce{X|E,r} \<wf>$ ensure that reduce rules can be safely
performed without implicit renaming.

Well-formedness of the reduce configurations ensures that the reduce transitions
require no implicit bound-variable renaming to preserve hygiene.
Well-formedness of the need configuration guarantees that it cannot be
``stuck'': since $x \in CV(E)$, a well-formed need configuration always has a
binder frame $[(\lambda x.[])\;t]$ that matches the variable under focus, so
the configuration can transition.

Well-formedness of configurations combined with rule $D.2$'s name
management ensures that machine evaluation respects variable binding
structure.
\begin{thm}
  If $t$ is a closed term of the calculus, then 
  $ \refocus{\emptyset | [\;],t} \<wf>$.
\end{thm}
\begin{proof}
  $\emptyset | [\;] \<wf>$ and $FV(t) \subseteq CV([\;]) = \emptyset.$
\end{proof}

\begin{thm}
  Let $C_1$ and $C_2$ be configurations.
  If $C_1 \<wf>$ and $C_1 \nam\ C_2$
  then $C_2 \<wf>$.
\end{thm}
\begin{proof}
  By cases on $\nam\ $. The cases are immediate except 
  when $C_1 \nam\ C_2$ by a reduce rule.  
  The proofs for both kinds of reduce configurations are similar, so we 
  present only one of the cases:
  \begin{case}[$C_1 = \reduce{X | E_1,(\kappa x.E_2)\;
      \answer{E_3,v}}$]\mbox{}\\
    By definition, $C_1 \nam\ C_2 = \rebuild{X | E, v}$, where $E = E_1 \comp
    E_3 \comp [(\lambda x.[])\; v] \comp E_2$.  Since the transition rule
    introduces no new active variables, $X$ should remain the
    same. Furthermore, all the active variables remain distinct, though $x$ is
    now introduced by the binder frame $[(\lambda x.[])\; v]$ rather than the
    cont frame $[(\kappa x.E_2)\;[]]$.  It remains to show that $FV(v)
    \subseteq CV(E)$ and that $FV(E) = \emptyset$.

    First, since $C_1 \<wf>$, it follows by inversion that 
    $X | (E_1 \comp [(\kappa x.E_2)\;[]] \comp E_3 ) \<wf>$
    and
    \begin{displaymath}
      FV(v) \subseteq CV(E_1 \comp [(\kappa x.E_2)\;[]] \comp E_3) = 
      CV(E_1 \comp E_3) \subseteq CV(E).
    \end{displaymath}
    By Lemma~\ref{lm:fvcv}, %
    \begin{math}
      FV(E_1) = FV(E_1 \comp [(\kappa x.E_2)\;[]]) = 
      FV(E_1 \comp [(\kappa x.E_2)\;[]] \comp E_3) = \emptyset.
    \end{math}

    Since $CV([(\kappa x.E_2)\;[]]) = \emptyset$, it follows from 
    Lemma~\ref{lm:fvcv} that
    $FV(E_3) \subseteq CV(E_1)$ and from the definition of $FV$ that $FV(E_2)
    \subseteq CV(E_1) \union \set{x}$.
    From these it follows that $FV(E)=\emptyset$.
  \end{case}

\end{proof}

In short, well-formedness of the reduce configurations ensures that the reduce
rules can be safely performed without any implicit renaming. Since the machine
preserves well-formedness, this property persists throughout evaluation.
The rest of this paper only considers well-formed configurations.

\subsection{An Abstract Machine for Call-by-need}
\label{sec:machine}

Putting together our observations from the previous section, we now present the
specification of the abstract machine.  Figure~\ref{fig:transitions} presents
its $\nam$ transitions rules.  We have derived a heap-less abstract machine for
call-by-need evaluation.
It replaces the traditional manipulation of a heap using store-based effects
with disciplined management of the evaluation stack using control-based
effects.  In short, state is replaced with control.

Machine evaluation of a program $t$ begins with $\refocus{\emptyset|[\;],t}$ and
terminates at $\Braket{X|\answer{E,v}}$.

\begin{figure*}
  \centering\small
  \begin{boxedarray}{@{}l@{}}
    \vspace{1ex}
   \boxed{\reduce{X|E,r}}\quad\desc{Reduce} \\
    \begin{ltransitionrule}
      (D.1) &
      \reduce{X | E_1, (\kappa x.E_2)\;\answer{E_3,v}} &
      \nam &
      \rebuild{X | E_1 \comp E_3 \comp [(\lambda x.[])\; v] \comp E_2, v} & \\[0.5ex]

      (D.2) &
      \reduce{X | E_1, \answer{E_2,\lambda x.t_1} \; t_2} &
      \nam &
      \refocus{X,x' | E_1 \comp E_2 \comp [(\lambda x'.[])\;t_2], t_1[x'/x]}
      & x' \notin X \\
    \end{ltransitionrule} \\  \\

    \vspace{1ex}
    \boxed{\refocus{X|E,t}}\quad\desc{Refocus} \\
    \begin{ltransitionrule}
      (F.1) &
      \refocus{X | E, x} & \nam & \need{X | E, x} & \\[0.5ex]

      (F.2) &
      \refocus{X | E, \lambda x.t} & \nam &
      \rebuild{X | E, \lambda x.t} & \\[0.5ex]

      (F.3) &
      \refocus{X | E, t_1\;t_2} & \nam &
      \refocus{X | E \comp [[]\;t_2], t_1} &
    \end{ltransitionrule} \\  \\

    \vspace{1ex}
    \boxed{\rebuild{X|E,v}}\quad\desc{Rebuild} \\
    \begin{ltransitionrule}
      (B.1) &
      \rebuild{X | E_b,v} & \nam & \Braket{X|\answer{E_b,v}} & \\[0.5ex]

      (B.2) &
      \rebuild{X | E_1 \comp [[]\; t] \comp E_b, v} & \nam &
      \reduce{X | E_1, \answer{E_b,v}\; t} & \\[0.5ex]
      
      (B.3) &
      \rebuild{X | E_1 \comp [(\kappa x.E_2)\; []]\comp E_b, v} & \nam &
      \reduce{X | E_1, (\kappa x.E_2)\; \answer{E_b,v}} & \\[0.5ex]
      \multicolumn{5}{l}{\text{where }E_b = \overline{[(\lambda x_i.[])\;t_i]}}
    \end{ltransitionrule} \\  \\

    \vspace{1ex}
    \boxed{\need{X|E,x}}\quad\desc{Need} \\
    \begin{ltransitionrule}
      (N.1) &
      \need{X | E_1 \comp [(\lambda x.[])\; t] \comp E_2, x} & \nam &
      \refocus{X | E_1 \comp [(\kappa x.E_2)\; []], t}  & \\[0.5ex]
      \multicolumn{5}{l}{\text{where }\lbrack(\lambda x.[])\; t] \notin E_2}
    \end{ltransitionrule}
 \end{boxedarray}
   \caption{Call-by-need Machine}
  \label{fig:transitions}
\end{figure*}

\section{Correctness of the Machine}

The previous section proves that the machine manipulates terms in a manner that
preserves variable binding. In this section, we prove that those manipulations
correspond to standard-order call-by-need evaluation.  

To proceed, we first establish correspondences between abstract machine
configurations and call-by-need terms.  As we have alluded to previously, 
abstract machine contexts correspond directly to calculus contexts:
\begin{displaymath}
\begin{boxedarray}{lcl}
    \mtoc |[ [\;] |] & = & [] \\[0.5ex]
    \mtoc |[ [[]\; t] \comp E |] & = & \mtoc |[E|]\; t \\[0.5ex]
    \mtoc |[ [(\kappa x.E_1)\;[]] \comp E_2 |] & = &
    (\lambda x.\mtoc |[E_1|][x])\; \mtoc |[E_2|] \\[0.5ex]
    \mtoc |[ [(\lambda x.[])\; t] \comp E |] & = &
    (\lambda x.\mtoc |[E|])\; t \\[0.5ex]
\end{boxedarray}
\end{displaymath}
Redexes also map to call-by-need terms:
\begin{displaymath}
\begin{boxedarray}{lcl}
    \mtoc |[ \answer{E,v}\;t |]  & = & (\mtoc |[E|][v])\;t \\[0.5ex]
    \mtoc |[ (\kappa x.E_1)\; \answer{E_2,v} |]  & = &
    (\lambda x.\mtoc |[E_1|][x])\;(\mtoc |[E_2|][v])
\end{boxedarray}
\end{displaymath}
Given that terms map identically to terms, configuration mapping is defined
uniformly:
\begin{displaymath}
\begin{boxedarray}{lcl}
    \mtoc |[ \Braket{X|E,?} |]  & = & \mtoc |[E|][\;\mtoc |[?|]\;]
\end{boxedarray}
\end{displaymath}
Since the calculus is defined over alpha equivalence classes, we reason
up to alpha equivalence when relating terms to machine configurations.

We now state our fundamental correctness theorems.
First we guarantee soundness, the property that every step of the abstract
machine respects standard-order reduction.

\begin{thm}
  If $t_1 = \mtoc|[C_1|]$ and $C_1 \nam\ C_2$, then
  $t_1 \Sr\; t_2$, for some $t_2 = \mtoc |[C_2|]$.
\end{thm}
\begin{proof}
  By cases on $\nam\ $. Only rules $D.1$ and $D.2$ are not immediate.  The other
  rules preserve equality under $\mtoc |[C|]$.
\end{proof}

\begin{corollary}[Soundness]\mbox{}\\
  If $t = \mtoc|[C|]$ and $C \Nam \Braket{X|\answer{E,v}}$, then
  $t \Sr\; a$, for some $a = \mtoc|[\Braket{X|\answer{E,v}}|]$.
\end{corollary}
\begin{proof}
  By induction on the length of the $\Nam$ sequence. 
\end{proof}

We also prove completeness, namely that abstract machine reduction subsumes
standard order reduction.  

\begin{thm}[Completeness]\mbox{}\\
  If $t = \mtoc|[C|]$ and $t \Sr\; a$, then $C \Nam \Braket{X|\answer{E,v}}$,
  with $a = \mtoc|[\Braket{X|\answer{E,v}}|]$.
\end{thm}
\begin{proof}
  This proof proceeds by induction on the length of $\Sr$ sequences.
  It utilizes Proposition~\ref{prp:fast-sr} to accelerate the $\sr$ rules
  in accordance with $\nam$.  It also relies on a number of lemmas to
  establish that $\nam$ will find the unique redex of a term from any
  decomposition of a term into a context $E$ and a subterm $t$.
\end{proof}

\begin{thm}[Correctness]
  If $t = \mtoc|[C|]$, then 
\[t \Sr\; a\quad\hbox{if and only if}\quad C \Nam
  \Braket{X|\answer{E,v}}
\]
with $a = \mtoc|[\Braket{X|\answer{E,v}}|]$.
\end{thm}

\subsection{Discussion}

This abstract machine has nice inductive properties.  The refocus rules always
dispatch on the outermost term constructor.  The rebuild and need rules
dispatch on a prefix of the context, though each has different criteria
for bounding the prefix.

The abstract machine's evaluation steps should not be seen as merely a
desperate search for a redex.  Rather, the machine exposes the fine-grain
structure of call-by-need evaluation, just as the CK machine and the Krivine
machine~\cite{krivine07machine} model evaluation for call-by-value and
call-by-name respectively.  Answers are the partial results of computations,
and the rebuild rules represent the process of reconstructing and
returning a result to a reduction site.  Furthermore, the need rules can be
viewed as a novel form of variable-lookup combined with lazy evaluation.
The evaluation context captures the rest of computation, but not in order:
variable references cause evaluation to skip around in a manner that is
difficult to predict. 

The way that variables behave in these semantics reveals a connection to
coroutines.  The reduction rule $D.2$ binds a variable to a delayed
computation; referencing that variable suspends the current computation and
jumps to its associated delayed computation.  Upon completion of that
computation, any newly delayed computations (i.e. binder frames) are added to
the evaluation context and the original computation is resumed.

The standard-order reduction relation of the call-by-need lambda calculus
defines an evaluator concisely but abstractly.  Surely unique decomposition,
standardization, and hygiene ensure the existence of a deterministic evaluator,
but these properties do not spell out the details or implications.  Based on a
reasoned inspection of standard-order reduction, we expose its computational
behavior and capture it in a novel abstract machine that has no store.  The
improvements to the initial machine produce a variant that effectively
assimilates computational information, explicitly accounts for variable hygiene
and thereby reveals coarse-grained operational structure implicit in
call-by-need standard-order evaluation.

\subsection{Extensions}

The machine presented above describes evaluation only for the pure lambda
calculus.  In this subsection, we introduce some features that are typical of
pragmatic programming languages.

\subsubsection{Let binding}

To help with a proof that the call-by-need calculus can simulate call-by-name,
Ariola and Felleisen introduce a \emph{let}-based calculus. The let-calculus
adds the construction $\mathrm{let}\;x=t\;\mathrm{in}\;t$ to the set of terms
and considers a new axiom:
\begin{displaymath}
      (\lambda x.t_1)\;t_2 \:= \: \mathrm{let}\; x=t_2\; \mathrm{in}\; t_1
\end{displaymath}
The let-calculus is formulated by taking this axiom as a reduction
rule running from left to right, and reformulating the original three axioms of
the call-by-need calculus in terms of $\mathrm{let}$ expressions.  These new
axioms can be justified by proving that they are derivable from the original
three axioms and the new $\mathrm{let}$ axiom.

One approach to producing an abstract machine that supports $\mathrm{let}$ is
to repeat the entire process described in this section, but focusing on the
let-calculus instead of the lambda calculus.  However, let-binding can
be retro-fitted to the current lambda calculus much more simply by reading the
new $\mathrm{let}$ axiom as a \emph{right-to-left} reduction rule:
\begin{displaymath}
     \mathrm{let}\; x=t_2\; \mathrm{in}\; t_1  \notion{} (\lambda x.t_1)\;t_2 
\end{displaymath}
Essentially, this rule indicates that $\mathrm{let}$ bindings in the calculus
can be understood as equivalent to an immediate application of an abstraction
to a term.  

The consequences for the na\"ive machine are the addition of a new kind of
redex:
\begin{displaymath}
 r ::= \dots | \mathrm{let}\; x=t_2\; \mathrm{in}\; t_1 
\end{displaymath}
and a new refocus rule:
\begin{displaymath}
  \refocus{E,\mathrm{let}\; x=t_2\; \mathrm{in}\; t_1}
  \mapsto
  \reduce{E,\mathrm{let}\; x=t_2\; \mathrm{in}\; t_1}
\end{displaymath}
and a new reduce rule:
\begin{displaymath}
  \reduce{E,\mathrm{let}\; x=t_2\; \mathrm{in}\; t_1}
  \mapsto
  \refocus{E,(\lambda x.t_1)\;t_2}
\end{displaymath}
Then, in the process of improving our machine, the reduce rule can be
fast-forwarded to reduce the introduced abstraction and application
immediately:
\begin{displaymath}
  \reduce{X|E,\mathrm{let}\; x=t_2\; \mathrm{in}\; t_1}
  \mapsto
  \refocus{X,x'|E\comp[(\lambda x'.[])\;t_2],t_1[x'/x]}
\end{displaymath}

This process confirms that the $\mathrm{let}$ expression form can be
comfortably treated as a conservative extension of the call-by-need lambda
calculus.  Its addition does not force a radical reconstruction of the abstract
machine.  As we show in Section~\ref{sec:circularity}, adding a circular
$\mathrm{letrec}$ construct to the language fundamentally alters the system.

\subsubsection{Constants}

\citet{plotkin75byname} augments the lambda calculus with two sets of
constants, the basic constants $b$ and the function constants $f$.
In that language, the basic constants are observables and serve as placeholders
for real programming language constants like numbers, strings, and so on.  The
function constants serve as placeholders for real primitive functions,
generally over basic constants.  The function constants are not first-class
values in the language, but instead appear as operators in primitive function
expressions.
\begin{displaymath}
\begin{boxedarray}{lcl}
  t & \produce & \dots | b | f\;t \\
  v & \produce & \dots | b
\end{boxedarray}  
\end{displaymath}
Function expressions always take a single term argument.  

The calculus is parametrized on a partial function $\delta$ that maps
function-value pairs to values.  As is standard, the notions of reduction for
the calculus are augmented to handle function constant applications.
\begin{displaymath}
  \begin{boxedarray}{lclr}
    f\;v_1 & \notion{} & v_2 & \text{if } \delta(f,v_1) = v_2
  \end{boxedarray}
\end{displaymath}
Since the notions of reduction for function constants require them to be
applied to values, the calculus must account for answers. 
To handle this, the calculus can be extended with a rule to 
commute answer bindings with function expressions, as is done for application
expressions:
\begin{displaymath}
  \begin{boxedarray}{lcl}
    f\;((\lambda x.a)\;t) & \notion{} & (\lambda x. f\;a)\;t
  \end{boxedarray}
\end{displaymath}
If function expressions were only defined for basic constant arguments, then
answer bindings could be discarded instead of commuted.  However this approach
imposes an ad hoc limitation on the semantics.  It would be cleaner and more
orthogonal to uniformly handle garbage collection with a specific notion of
reduction~\cite{ariola97need}.

Function constants must be considered in the evaluation contexts.
A function constant must force its argument in order to produce a value, so
upon encountering a function constant application, evaluation must continue in
its argument:
\begin{displaymath}
\begin{boxedarray}{lcl}
  E & \produce & \dots | f\;E
\end{boxedarray}
\end{displaymath}

The other notions of reduction remain exactly the same, but since basic
constants are values, they are subject to rules that manipulate values and
answers. For example, if the set of
basic constants includes numbers and the set of function constants includes
operations on numbers, then by the deref rule:
\begin{displaymath}
  (\lambda x . \mathrm{add1}\;x)\; 5 \notion\ (\lambda x.\mathrm{add1}\;5)\;5
\end{displaymath}

The abstract machine requires few changes to accommodate these additions.
The set of redexes is extended to include function expressions:
\begin{displaymath}
  r ::= \dots | f\;a
\end{displaymath}
The refocus rules are generalized to create context frames for function
constant applications and to rebuild when any value, lambda abstraction
or basic constant, is encountered:
\begin{displaymath}
  \begin{array}{rcl}
    \refocus{X|E,f\;t} & \nam & \refocus{X|E\comp [f\;[]],t} \\   
    \refocus{X|E,v} & \nam & \rebuild{X|E,v}   
  \end{array}
\end{displaymath}
Furthermore, rebuild must account for function constant applications:
\begin{displaymath}
  \refocus{X|E \comp [f\;[]] \comp E_b,v} \nam\ \reduce{X|E,f\;\answer{E_b,v}}
\end{displaymath}
Finally, the reduce rule for $f\;a$, pushes binder frames upwards and appeals
to the delta rule $\delta(f,v)$ for its result:
\begin{displaymath}
   \reduce{X|E,f\;\answer{E_b,v}} \nam\ \rebuild{X|E\comp E_b,\delta(f,v)}
\end{displaymath}
Since $\delta(f,v)$ yields only values, the reduce configuration for $f\;a$ can
immediately transition to rebuild from this result.

\subsection{Lazy Constructors}

As pointed out by \citet{ariola97need}, the call-by-need lambda calculus can be
easily extended with support for lazy constructors.  The rules for constructors
and destructors can be inferred from the church encoding for pairs:
\begin{displaymath}
  \begin{boxedarray}{rcl}
  \mathtt{cons} & \equiv & \lambda x_1.\lambda x_2.\lambda d.d\;x_1\;x_2 \\
  \mathtt{car} & \equiv & \lambda p.p\;(\lambda x_1.\lambda x_2.x_1) \\
  \mathtt{cdr} & \equiv & \lambda p.p\;(\lambda x_1.\lambda x_2.x_2) \\
  \end{boxedarray}
\end{displaymath}

To add support for lazy pairs, we first extend the syntax of the language:
\begin{displaymath}
  \begin{boxedarray}{rcl}   
    t & \produce & \dots | \mathtt{cons}\;t\;t | \mathtt{car}\;t |
    \mathtt{cdr}\;t | \consd{x}{x} \\
    v & \produce & \dots | \consd{x}{x} \\
    E & \produce & \dots | \mathtt{car}\;E | \mathtt{cdr}\;E
  \end{boxedarray}
\end{displaymath}
The $\mathtt{cons}\;t_1\;t_2$ expression creates a lazy pair of its two
arguments, $t_1$ and $t_2$.  The $\mathtt{car}\;t$ and $\mathtt{cdr}\;t$
expressions evaluate their respective arguments and extract the first or second
element of the resulting pair.  As with function constants in the previous
section, the lazy pair constructor and destructors are second-class. To express
first-class constructors, these may be eta-expanded into, e.g., $\lambda
x_1.\lambda x_2.\mathtt{cons}\;x_1\;x_2$.  The $\consd{x_1}{x_2}$ value is a
representation of a lazy pair.  It contains variables that refer to shared
computations.

According to the evaluation contexts, evaluation may focus on the argument of a
$\mathtt{car}$ or $\mathtt{cdr}$ expression.  However, evaluation never
directly operates on the arguments to a constructor.  This is why the
evaluation contexts for the language do not examine the arguments to
$\mathtt{cons}$.  As a direct consequence, the standard call-by-need reduction
rules will never substitute a value for a variable inside a pair.  A variable
inside a pair can only be evaluated after decomposing the pair using
$\mathtt{car}$ or $\mathtt{cdr}$.

The Church encoding of pairs motivates the following notions of reduction:
\begin{displaymath}
  \begin{boxedarray}{rcl}    
    \mathtt{cons}\;t_1\;t_2 & \notion{}
    & (\lambda x_1.(\lambda x_2.\consd{x_1}{x_2})\;t_2)\;t_1 \\
    \mathtt{car}\;\consd{x_1}{x_2} & \notion{} & x_1 \\
    \mathtt{cdr}\;\consd{x_1}{x_2} & \notion{} & x_2 \\
  \end{boxedarray}
\end{displaymath}
These rules model $\mathtt{cons}$ as it is applied to two arguments. The
$\mathtt{car}$ and $\mathtt{cdr}$ operations each expose a previously
inaccessible variable reference to evaluation.

To accommodate lazy pairs in the abstract machine,
we extend the set of redexes:
\begin{displaymath}
  r ::= \dots | \mathtt{cons}\;t_1\;t_2 | \mathtt{car}\;a | \mathtt{cdr}\;a
\end{displaymath}
and we introduce several new rules.
The first set of rules extends the refocus stage of the machine to handle our
extensions. 
\begin{displaymath}
  \begin{boxedarray}{rcl}    
  \label{nam:cons1}
  \refocus{X|E,\mathtt{cons}\;t_1\;t_2} & \nam &
  \reduce{X|E,\mathtt{cons}\;t_1\;t_2} \\

  \refocus{X|E,\mathtt{car}\;t} & \nam & 
  \refocus{X|E \comp [\mathtt{car}\;[]], t}  \\

  \refocus{X|E,\mathtt{cdr}\;t} & \nam & 
  \refocus{X|E \comp [\mathtt{cdr}\;[]], t} \\

  \end{boxedarray}
\end{displaymath}
A $\mathtt{cons}$ expression is immediately ready for reduction. The
$\mathtt{car}$ and $\mathtt{cdr}$ expressions proceed to evaluate their
respective arguments.
Another set of rules returns an answer to $\mathtt{car}$ or $\mathtt{cdr}$.
\begin{displaymath}
  \begin{boxedarray}{rcl}    
  \label{nam:cons2}

  \rebuild{X| E \comp [\mathtt{car}\;[]] \comp E_b,v} & \nam &
  \reduce{X|E,\mathtt{car}\;\answer{E_b,v}} \\

  \rebuild{X| E \comp [\mathtt{cdr}\;[]] \comp E_b,v} & \nam &
  \reduce{X|E,\mathtt{cdr}\;\answer{E_b,v}} \\
  \end{boxedarray}
\end{displaymath}
Finally, a set of rules corresponds to the notions of reduction. 
\begin{displaymath}
  \begin{boxedarray}{rcl}    
  \label{nam:cons3}
  \reduce{X|E,\mathtt{cons}\;t_1\;t_2} & \nam &
  \rebuild{X,x_1,x_2|
    E\comp [(\lambda x_1.[])\;t_1]\comp [(\lambda x_2.[])\;t_2],
    \consd{x_1}{x_2}} \\

  \reduce{X|E_1,\mathtt{car}\;\answer{E_2,\consd{x_1}{x_2}}} & \nam &
  \need{X|E_1 \comp E_2, x_1} \\

  \reduce{X|E_2,\mathtt{cdr}\;\answer{E_2,\consd{x_1}{x_2}}} & \nam & 
    \need{X|E_1 \comp E_2, x_2} \\
  \end{boxedarray}
\end{displaymath}
Lazy pair construction creates new binder frames for the two terms and produces
a pair that references them.  To evaluate a destructor, the binder frames
associated with the answer are pushed upwards and the corresponding variable
reference is extracted from the pair and the value of its associated
computation is demanded.

\subsection{Circularity}
\label{sec:circularity}
As pointed out by \citet{ariola95need}, the presence of constructors makes it
possible to observe duplicated constructors when the $\mathsf{Y}$ combinator is
used to express recursion.

Consider the expression:
\begin{displaymath}
  Y(\lambda y.\mathtt{cons}\;1\;y) \equiv  
  (\lambda f.(\lambda x.f\;(x\;x))\;(\lambda x.f\;(x\;x)))\; 
  (\lambda y.\mathtt{cons}\;1\;y)
\end{displaymath}

Standard-order reduction in the calculus proceeds as follows, underlining
either the active variable reference or the current redex:
\begin{align*}
 &  (\lambda f.(\lambda x.\underline{f}\;(x\;x))\;(\lambda x.f\;(x\;x)))\; 
  (\lambda y.\mathtt{cons}\;1\;y) \\
 \notion\enspace\ & (\lambda f.(\lambda x.(\lambda y.\underline{\mathtt{cons}\;1\;y})\;
  (x\;x))\;
  (\lambda x.f\;(x\;x)))\; 
  (\lambda y.\mathtt{cons}\;1\;y) \\
 \notion\enspace\ &   (\lambda f.(\lambda x.(\lambda y.(\lambda x_1.
  (\lambda x_2.\consd{x_1}{x_2})\;y)\;1)\;
  (x\;x))\;(\lambda x.f\;(x\;x)))\; 
  (\lambda y.\mathtt{cons}\;1\;y) 
\end{align*}
Because of the two references to $f$ in $\mathsf{Y}$, the term $(\lambda
y.\mathtt{cons}\;1\;y)$ is copied and ultimately recomputed each time the term
is needed.  Ideally a recursive $\consd{x}{x}$ value could refer to itself and
not recompute its value whenever its $\mathtt{cdr}$ is demanded.  Duplicated
computation does not correspond to how recursion and lazy constructors interact
in typical semantics for lazy evaluation~\cite{henderson76lazy}. An
implementation would create a single self-referencing cons cell.

Because of the scoping rules for $\lambda$, there is no way to explicitly
define truly circular structures in this calculus.  To address this, Ariola and
Felleisen introduce a \emph{letrec}-based calculus, inspired by the circular
calculus of \citet{ariola94rewriting,ariola97recursion}.  The
syntax of the letrec-calculus follows:
\begin{displaymath}
\begin{boxedarray}{lcl}
  t & \produce & x | \lambda x.t | t\;t |
  \<letrec> D \<lin> t \\
  D & \produce & x_1 \<be> t_1, \dots, x_n \<be> t_n \\
  v & \produce & \lambda x.t \\
  a & \produce & v | \<letrec> D \<lin> a \\
  E & \produce & [] | E\; t | \<letrec> D \<lin> E  \\
  & | &\<letrec> D, x \<be> E \<lin> E[x] \\
  & | &\<letrec> x_n \<be> E, D[x,x_n] \<lin> E[x] \\
  D[x,x_n] & \produce & x \<be> E[x_1], x_1 \<be> E[x_2], \dots, 
  x_{n-1} \<be> E[x_n], D
\end{boxedarray}  
\end{displaymath}
The letrec-calculus is similar to the let-calculus in how it adds an explicit
binding form, but each $\<letrec>\!\!$ expression can contain an unordered set
$D$ of mutually recursive bindings $x \<be> t$ for distinct variables.

The evaluation contexts for the letrec-calculus resemble those for
the let-calculus.  The contexts
\begin{math}
\<letrec> D, x \<be> E \<lin> E[x] 
\end{math}
and
\begin{math}
\<letrec> x_n \<be> E, D[x,x_n] \<lin> E[x]
\end{math}
represent evaluation taking place in the binding position of a $\<letrec>\!\!$
expression.  The first form expresses that a variable has been referenced in
the body of the $\<letrec>\!\!$ and that variable's definition is currently
under evaluation.  The second form expresses that some variables bound in the
$\<letrec>\!\!$ form depend on each other, and a variable reference in the body
of the letrec has forced evaluation of this chain of variable references. This
chain of dependencies is denoted by the syntax $D[x,x_n]$.  The definition of
the last referenced variable in the chain of dependencies, $x_n$, is currently
being evaluated.  For evaluation to proceed, all the variables in a chain of
dependencies must be disjoint.  The letrec calculus regards a cyclic dependency
chain as a diverging computation, a stuck expression that is not a valid
answer.

The notions of reduction for the letrec calculus follow:
\begin{displaymath}
\begin{boxedarray}{rlcl@{}}
(1)&(\lambda x.t_1)\;t_2 & \notion & \<letrec> x \<be> t_2 \<lin> t_1 \\
(2)&\<letrec> D, x \<be> v \<lin> E[x]  & \notion & 
\<letrec> D, x \<be> v, \<lin> E[v] \\
(3)&\<letrec> x_n \<be> v, D[x,x_n] \<lin> E[x]
 & \notion &  
\<letrec> x_n \<be> v, D[x,v] \<lin> E[x] \\
(4)&(\<letrec> D \<lin> a)\;t & \notion & \<letrec> D \<lin> a\;t \\

(5)&\begin{array}[t]{@{}l@{}}
\<letrec> 
x_{n} \<be> (\<letrec> D \<lin> a), D[x,x_n]\\
\<lin> E[x]
\end{array}
 & \notion &  
\begin{array}{@{}l@{}}
\<letrec> D, x_{n} \<be> a, D[x,x_n] \<lin> E[x]
\end{array}
\\

(6) &
\<letrec> D_1, x \<be> (\<letrec> D_2 \<lin> a)\<lin> E[x]
& \notion &  
\begin{array}{@{}l@{}}
\<letrec> D_2, D_1, x \<be> a \<lin> E[x]
\end{array}

\end{boxedarray}  
\end{displaymath}
The rules operate as follows.
Rule (1) is analogous to the equivalent let-calculus rule.
Rule (2) is analogous to the basic dereference rule.
Rule (3) resolves the last variable reference in a chain of dependency.
The syntax $D[x,v]$ expresses  replacing $x_{n-1} \<be> E[x_n]$ with 
$x_{n-1} \<be> E[v]$ in the list of bindings. 
Rules (4) through (6) are associativity rules that percolate bindings
upward to ensure proper sharing.  
Rules (5) and (6) are important for recursion. They lift recursive bindings
from the definition of a letrec-bound variable $x$ and incorporate them into
the same $\<letrec>\!\!$ expression that binds $x$.  In essence they expand the
set of variables bound by the outer $\<letrec>\!\!$ expression.

Using the letrec-calculus, the corresponding example using \texttt{cons} is as
follows:
\begin{align*}
&  
\<letrec> y \<be> \mathtt{cons}\;1\;y \<lin> y \\
\notion\enspace\ &
\<letrec> y \<be> (\<letrec> x_1 \<be> 1 \<lin> 
(\<letrec> x_2 \<be> y \<lin> 
\consd{x_1}{x_2})) \<lin> y \\
\notion\enspace\ &
\<letrec> x_1 \<be> 1, y \<be> 
(\<letrec> x_2 \<be> y \<lin> 
\consd{x_1}{x_2}) \<lin> y \\
\notion\enspace\ &
\<letrec> x_1 \<be> 1, x_2 \<be> y, y \<be>
\consd{x_1}{x_2} \<lin> y \\
\notion\enspace\ &
\<letrec> x_1 \<be> 1, x_2 \<be> y, y \<be> 
\consd{x_1}{x_2} \<lin> \consd{x_1}{x_2}
\end{align*}
In this reduction, the recursive cons cell has been computed once and for all.
No reference to the original \texttt{cons} lazy constructor remains.

\subsubsection{The letrec machine}

Now we express the call-by-need letrec-calculus as an abstract machine.
Development of this machine proceeds along the same lines as the non-circular
machine.

As described in the calculus,
$\<letrec>\!\!$ bindings need not be ordered, but dependency chains have a
natural order imposed by the order of dependencies.  Furthermore, cyclic
dependency chains are provable divergences in the letrec-calculus.  In the face
of such a cycle, a program is no longer subject to reduction.  We can model
this by letting the machine get ``stuck''.

The basic terms for the machine remain close to those of the calculus:
\begin{displaymath}
\begin{boxedarray}{lcl}
  t & \produce & x | \lambda x.t | t\;t | \<letrec> D^{+} \<lin> t \\[0.5ex]
  D & \produce & \seq{x_i \<be> t_i} \\
  v & \produce & \lambda x.t \\
\end{boxedarray}  
\end{displaymath}
As with the calculus, $D$ refers to sets of recursive bindings.  As needed,
we distinguish  possibly empty sets of bindings, $D^{*}$, from nonempty
sets $D^{+}$.  This precision is needed to discuss machine behavior.
The values of this machine are the lambda abstractions, but now answers are
defined as values wrapped in zero or more tiers of $\<letrec>\!\!$ bindings.

As we did for the prior machine, we introduce some representation changes that
help with presenting the letrec-machine.  The evaluation context is once again
replaced with a list of context frames.
\begin{displaymath}
\begin{boxedarray}{lcl}
  \PP{x_n}{x_0} & \produce & 
\left\{\braket{x_n,E_n}::\braket{x_{n-1},E_{n-1}}::\cdots::\braket{x_0,E_0}\right\} \\
  f & \produce & []\; t | \LF D^{+}  \<lin> []  \\
  & | &\LF x \<be> [], D^{*} \<lin> E \\
  & | &\LF x_{n+1} \<be> [], \PP{x_n}{x_0}, D^{*} \<lin> E \\

  E & ::= & [\;] | [f] \comp E | E \comp [f] \\[0.5ex]
  & &\text{where } E \comp [\;] = [\;] \comp E = E\\[0.5ex]
  & &\text{and } E_1 \comp (E_2\comp E_3) = (E_1\comp E_2)\comp E_3\\  
\end{boxedarray}
\end{displaymath}
The operand frame $[[]\;t]$ and binding frame $[\LF D^{+} \<lin> []]$ are directly
analogous to the corresponding frames in the original abstract machine.  The
original cont frame, on the other hand, splits into two variants.  
The frame 
$[\LF x_{n+1} \<be> [], \PP{x_0}{x_n}, D^{*} \<lin> E[x_0]]$
captures chain of dependencies that is currently under evaluation.  The
expression $D^{*}$ stands for the inactive bindings in the frame, while the
expression $\PP{x_0}{x_n}$ indicates a chain of dependencies.  While no
particular ordering is imposed on the bindings in $D^{*}$, the chain of
dependencies represented by $\PP{x_n}{x_0}$ is ordered. 
Our machine representation separates dependencies from other bindings.  In the
calculus, $D[x_0,x_n]$ also includes any other bindings $D$.  In the machine,
$\PP{x_n}{x_0}$ only captures the dependencies, and the other bindings $D^{*}$
are explicitly indicated.
Machine dependencies
$\PP{x_n}{x_0} \equiv
\left\{\braket{x_n,E_n}::\braket{x_{n-1},E_{n-1}}::\cdots::
  \braket{x_0,E_0}\right\}$
correspond to calculus dependencies $D[x_0,x_n] \equiv x \<be> E[x_1], x_1
\<be> E[x_2], \dots, x_{n-1} \<be> E[x_n]$.  We allow $\PP{x_{-1}}{x_0}$ to
denote an empty chain.  Observe that the order of the
machine dependencies is reversed.  This ordering expresses that dependencies
are resolved in last-in first-out order.  When the value of $x_{n+1}$ is
computed, its value will be used to compute the value of $x_{n}$ and so on
until the value of $x_0$ is computed and its value is returned to the body of
the $\<letrec>\!\!$ expression.  Machine dependencies form a stack of
computations that reference one another.  This behavior is clarified in the
behavior of the cyclic abstract machine.

The frame $[\LF x \<be> [], D^{*} \<lin> E]$ closely resembles the old cont
frame $[(\kappa x.E)\;[]]$, except that the $\LF$ form may bind other variables
as well.  The absence of $\PP{x_n}{x_0}$ in this frame explicitly indicates that
it has no chain of dependencies to be evaluated before substituting into the
body of the $\LF$.

Machine answers are still binding-value pairs, but now each binding
is a recursive binding of multiple variables.
\begin{displaymath}
\begin{array}{lclr}
  a & \produce & \answer{E,v} &
  \text{where }E = \seq{[\LF D^{+}_i \<lin> []]} \\
\end{array}
\end{displaymath}

The set of redexes for the abstract machine follows.
\begin{displaymath}
\begin{array}{lcl}
  r & \produce & a\;t \\ 
  & | & \LF x \<be> a, D^{*} \<lin> E \\
  & | & \LF x_{n+1} \<be> a, \PP{x_n}{x_0}, D^{*} \<lin> E \\
\end{array}
\end{displaymath}
The first redex form is the same as a form from the original machine.  The
second form is analogous to the $(\kappa x.E)\;a$ form from the original
machine.  The final redex form captures the case where one link in the chain of
dependencies is about to be resolved.

Figure~\ref{fig:letrec-machine} presents the transition rules of the cyclic
abstract machine. The machine relies on an operator $\mathcal{F}(E_b)$, which
given a list of binder frames $\seq{[\LF D^{+}_i \<lin> []]}$, flattens
them into a single binder frame $[\LF \seq{D^{+}_i} \<lin> []]$. This
operation captures in aggregate the treatment of bindings by rules $(5)$ and
$(6)$ of the calculus.

The first three refocus rules are the same as the lambda calculus, while the
fourth rule is analogous to the equivalent rule for the let-calculus.  
The rebuild rules are also analogous to the lambda calculus, though now the
binder frames have the form $[\LF D^{+}_i  \<lin> []]$.

The letrec-machine has two need rules, $(N.1)$ for a variable
reference in the body of the corresponding $\<letrec>$, and $(N.2)$
for a variable reference that extends a (possible empty) chain of
dependencies. The need rule does not consider variable bindings that are
currently in a dependency chain, so evaluation will get stuck upon arriving at
a cycle.

There are four reduce rules.  The $(D.1)$ rule, substitutes a value into the
body of a letrec after its value has been computed.  The $(D.2)$ rule resolves
the most recent reference in a chain of dependencies.  Having computed the
value of $x_{n+1}$, it returns to computing the value of $x_n$, which needed
$x_{n+1}$'s value.  If the chain of dependencies has only one element
(i.e. $\PP{x_0}{x_0}$), then the chain is fully resolved.  The $(D.3)$ rule
handles when an answer is applied to an expression. It combines calculus rules
$(4)$ and $(1)$.  Finally, the $(D.4)$ rule handles $\<letrec>$ occurrences in
the source program.  In order to address hygiene, this rule must simultaneously
substitute for every bound variable in each binding as it focuses on the body
of the $\<letrec>$ expression.

The following is a machine trace of the cyclic $\mathtt{cons}$ example:
\begin{displaymath}
  \begin{array}[]{lllll}
& & \refocus{\emptyset | [\;], \<letrec> y \<be> \mathtt{cons}\;1\;y \<lin> y} \\
& \mapsto & 
\reduce{\emptyset | [\;], \<letrec> y \<be> \mathtt{cons}\;1\;y \<lin> y} \\
& \mapsto & 
\refocus{\set{y} | [\LF y \<be> \mathtt{cons}\;1\;y \<lin> []], y} \\
& \mapsto &
\need{\set{y} | [\LF y \<be> \mathtt{cons}\;1\;y \<lin> []], y}    \\
& \mapsto &
\refocus{\set{y} | [\LF y \<be> [] \<lin> [\;]], \mathtt{cons}\;1\;y} \\    
& \mapsto &
\reduce{\set{y} | [\LF y \<be> [] \<lin> [\;]], \mathtt{cons}\;1\;y}  \\
& \mapsto &
\rebuild{\set{y,x_1,x_2}| [\LF y \<be> [] \<lin> [\;]] \comp
[\LF x_1 \<be> 1 \<lin> []] \comp [\LF x_2 \<be> y \<lin> []], 
\consd{x_1}{x_2}}  \\
& \mapsto &
\reduce{\set{y,x_1,x_2}| [\;], \LF y \<be> \answer{[\LF x_1 \<be> 1 \<lin> []] 
    \comp [\LF x_2 \<be> y \<lin> []],\consd{x_1}{x_2}} \<lin> [\;]}  \\
& \mapsto &
\rebuild{\set{y,x_1,x_2}| [\LF y \<be> \consd{x_1}{x_2}\!, x_1 \<be> 1,
  x_2 \<be> y \<lin> []], \consd{x_1}{x_2}}  \\
& \mapsto &
\braket{\set{y,x_1,x_2}| \answer{[\LF y \<be> \consd{x_1}{x_2}\!, x_1 \<be> 1,
  x_2 \<be> y \<lin> []], \consd{x_1}{x_2}}}  \\
\end{array}
\end{displaymath}

\section{Simulating Call-by-need Using Control}

As we allude to above, call-by-need machine evaluation is highly suggestive of
delimited control operations, but the connection is indirect and mixed with the
other details of lazy evaluation.  In this section, we directly interpret this
connection in the terminology of delimited control.

Based on the operational behavior of the abstract machine from
Figure~\ref{fig:transitions}, we derive a simulation of call-by-need execution
under call-by-value augmented with delimited control operators.  In particular,
we translate call-by-need terms into the framework
of~\citet{dybvig07monadic}. First we overview the language of delimited control
operations.  Then we describe how the abstract machine performs delimited
control operations.  Next we present the simulation of call-by-need using
delimited control.  Finally we show its correctness.

\begin{landscape}
\begin{figure*}
  \centering\small
  \begin{boxedarray}{@{}l@{}}
    \vspace{1ex}
   \boxed{\reduce{X|E,r}}\quad\desc{Reduce} \\
    \begin{ltransitionrule}
      (D.1) & 
      \reduce{X | E_1, 
        (\LF x \<be> \answer{E_2,v},D^{*} \<lin> E_3)} & \nam &
      \rebuild{X | E_1 \comp
        [\LF x \<be> v, \flatten{E_2},D^{*} \<lin> []]  \comp E_3,v} \\

      (D.2) &
      \reduce{X | E_1, 
        (\LF x_{n+1} \<be> \answer{E_2,v}, \PP{x_n}{x_0}, D^{*} \<lin> E_3)} 
      & \nam & 
      \rebuild{X | E_1 \comp
        [\LF x_n \<be> [], \PP{x_{n-1}}{x_0}, x_{n+1} \<be> v,
        \flatten{E_2},D^{*} \<lin> E_3]  \comp E_n,v} \\

      (D.3) &
      \reduce{X | E_1, \answer{E_2,\lambda x.t_1} \; t_2} &
      \nam &
      \refocus{X,x' | E_1 \comp E_2 \comp [\LF x' \<be> t_2 \<lin> []],
        t_1[x'/x]}
      \quad x' \notin X \\

      (D.4) & 
      \reduce{X | E, \<letrec> \seq{x_i \<be> t_i} \<lin> t} & \nam &
      \refocus{X,\seq{x'_i} | E \comp 
        [\LF \seq{x'_i \<be> t_i[\seq{x'_i/x_i}]} \<lin> []], t} 
      \quad \seq{x'_i} \cap X = \emptyset
    \end{ltransitionrule} \\  \\

    \vspace{1ex}
    \boxed{\refocus{X|E,t}}\quad\desc{Refocus} \\
    \begin{ltransitionrule}
      (F.1) &
      \refocus{X | E, x} & \nam & \need{X | E, x} & \\[0.5ex]

      (F.2) &
      \refocus{X | E, \lambda x.t} & \nam &
      \rebuild{X | E, \lambda x.t} & \\[0.5ex]

      (F.3) &
      \refocus{X | E, t_1\;t_2} & \nam &
      \refocus{X | E \comp [[]\;t_2], t_1} & \\

      (F.4) & 
      \refocus{X | E, \<letrec> D^{+} \<lin> t} & \nam &
      \reduce{X | E, \<letrec> D^{+} \<lin> t} \\
    \end{ltransitionrule} \\  \\

    \vspace{1ex}
    \boxed{\rebuild{X|E,v}}\quad\desc{Rebuild} \\
    \begin{ltransitionrule}
      (B.1) &
      \rebuild{X | E_b,v} & \nam & \Braket{X|\answer{E_b,v}} & \\[0.5ex]

      (B.2) &
      \rebuild{X | E_1 \comp [[]\; t] \comp E_b, v} & \nam &
      \reduce{X | E_1, \answer{E_b,v}\; t} & \\[0.5ex]

      (B.3) &
      \rebuild{X | E_1 \comp 
        [(\LF x \<be> [],D^{*} \<lin> E_2)] \comp E_b, v} & \nam &
      \reduce{X | E_1, 
        (\LF x \<be> \answer{E_b,v},D^{*} \<lin> E_2)} & \\

      (B.4) &
      \rebuild{X | E_1 \comp 
        [(\LF x_{n+1} \<be> [], \PP{x_n}{x_0}, D^{*} \<lin> E_2] \comp E_b, v} & \nam &
      \reduce{X | E_1, 
        (\LF x_{n+1} \<be> \answer{E_b,v}, \PP{x_n}{x_0}, D^{*} \<lin> E_2)} \\
      \multicolumn{5}{l}{\text{where }E_b = \overline{[\LF D^{+}_i  \<lin> []]}}
    \end{ltransitionrule} \\  \\

    \vspace{1ex}
    \boxed{\need{X|E,x}}\quad\desc{Need} \\
    \begin{ltransitionrule}
      (N.1) &
      \need{X | E_1 \comp 
        [\LF (x \<be> t, D^{*}) \<lin> []] \comp E_2^\dagger, x} & \nam &
      \refocus{X | E_1 \comp
        [\LF x \<be> [], D^{*} \<lin> E_2] , t}  & \\[0.5ex]

      (N.2) &
      \need{X | E_1 \comp 
        [\LF x_n \<be> [], \PP{x_{n-1}}{x_0}, (x \<be> t, D^{*}) 
        \<lin> E_2]  \comp E_n^\dagger, x} & \nam &
      \refocus{X | E_1 \comp
        [\LF x \<be> [], \PP{x_n}{x_0}, D^{*} \<lin> E_2] , t}  &
      \\[0.5ex]
      \multicolumn{5}{l}{\text{where } (x \<be> t) \notin E^\dagger}
    \end{ltransitionrule}
 \end{boxedarray}
   \caption{Letrec Machine}
  \label{fig:letrec-machine}
\end{figure*}
\end{landscape}

\subsection{Delimited Control Operators}
\citet{dybvig07monadic} define a language with delimited control operators.
We explain these operators using a simplified variant of
the defining machine semantics.
\begin{displaymath}
\begin{boxedarray}{lcl}
  t & \produce & x | v | t\;t | \<newPrompt> | \<pushPrompt> t\;t \\[0.5ex] 
  & | & \<withSubCont> t\; t | \<pushSubCont> t\; t \\[0.5ex]
  v & \produce & \lambda x.t | p | \Braket{M} \\[0.5ex]
  E & \produce & \hole | E[[]\;t] | E[(\lambda x.t)\;[]] 
  | E[\<pushPrompt> []\;t] \\[0.5ex]
  & | & E[\<withSubCont> []\; t] | E[\<withSubCont> p\; []] \\[0.5ex]
  & | & E[\<pushSubCont> []\; t]\\[0.5ex]
  M & \produce & [\;] | E : M  | p:M  \\[0.5ex] 
  p & \in & \mathbb{N} \\[0.5ex]
\end{boxedarray}  
\end{displaymath}
The language extends the call-by-value untyped lambda calculus with the four
operators $\<newPrompt>$, $\<pushPrompt>\!$, $\<withSubCont>\!$, and
$\<pushSubCont>\!$ as well as two new values: first-class \emph{prompts} $p$,
and first-class delimited continuations $\Braket{M}$.  Its control structure is
defined using evaluation contexts $E$, and metacontexts $M$, which are
lists that interleave prompts and contexts. 
Metacontexts use Haskell list notation.
Prompts are modeled using natural numbers.

A program state comprises an expression $t$, continuation $E$, metacontinuation
$M$, and fresh prompt source $p$.  The initial state for a
program $t$ is $\hole[t],[\;],0$.
\begin{displaymath}
\begin{boxedarray}{lcl}
      E[(\lambda x.t)\; v],M,p & \mapsto & E[t[v/x]],M,p \\[0.5ex]
      E[\<newPrompt>],M,p & \mapsto & E[p],M,p+1 \\[0.5ex]
      E[\<pushPrompt> p_1\;t],M,p_2 & \mapsto & \hole[t],p_1:E:M,p_2 \\[0.5ex]
  E[\<withSubCont> p_1\;\lambda x.t],M_1++(p_1:M_2),p_2 &\mapsto &
\hole[t[\Braket{E:M_1}/x]],M_2,p_2 \\[0.5ex]
\text{ where } p_1 \notin M_1 \\[0.5ex]
E[\<pushSubCont>\!\Braket{M_1}\;t],M_2,p & \mapsto & 
\hole[t],M_1 ++ (E:M_2),p \\[0.5ex]
      \hole[v],E:M,p & \mapsto & E[v],M,p \\[0.5ex]
      \hole[v],p_1:M,p_2 & \mapsto & \hole[v],M,p_2 \\[0.5ex]
\end{boxedarray}  
\end{displaymath}

The four operators manipulate delimited continuations, or
\emph{subcontinuations}, which are part of an execution context.  The
$\<withSubCont>\!$ operator takes a prompt and a function; it captures the
smallest subcontinuation that is delimited by the prompt and passes it to the
function.  The non-captured part of the continuation becomes the new
continuation.  The prompt instance that delimited the captured subcontinuation
is discarded: it appears in neither the captured subcontinuation nor the
current continuation.  This operator
generalizes~$\mathcal{F}$~\cite{felleisen88prompts} and
\texttt{shift}~\cite{danvy90abstracting}.

The $\<pushSubCont>\!$ operator takes a subcontinuation and an expression; it
composes the subcontinuation with the current continuation and proceeds to
evaluate its second argument in the newly extended continuation.

The $\<pushPrompt>\!$ operator takes a prompt and an expression; it extends the
current continuation with the prompt and evaluates the expression in the newly
extended continuation.  The $\<newPrompt>$ operator returns a distinguished
fresh prompt each time it is called. These two operators generalize the
delimiting operators $\#$~\cite{felleisen88prompts} and
\texttt{reset}~\cite{danvy90abstracting}, which extend a continuation with a
single common delimiter.

To illustrate these operators in action, we consider a program that uses
arithmetic and conditionals:
\begin{displaymath}
\begin{array}{@{}l@{}}
\<let> p = \<newPrompt>  \\
\<in> 2 + \<pushPrompt> p \\
\qquad\qquad
\begin{array}{@{}l@{}}
\<if> (\<withSubCont> p \\
\qquad (\lambda k.(\<pushSubCont> k\; \mathit{False}) +\\
\qquad\quad (\<pushSubCont> k\; \mathit{True}))) \\
\<then> 3 \\
\<else> 4 \\
\end{array}
\end{array}
\end{displaymath}
A fresh prompt is bound to $p$ and pushed onto the continuation just prior to
evaluation of the $\<if>\!$ expression.  $\<withSubCont>\!$ captures the
subcontinuation $[\<if> []\;\<then> 3\;\<else>4]$, which was delimited by~$p$,
and binds it to $k$.  The subcontinuation $k$ is pushed twice, given the value
$\mathit{False}$ the first time and $\mathit{True}$ the second.  The result of
evaluation is the expression $2 + 4 + 3$ which yields $9$.

\subsection{Delimited Control Na\"ively Simulates the Machine} 

The call-by-need abstract machine performs two different kinds of partial
control capture.  To review, the rebuild and need rules of the abstract machine
both capture some portion of the evaluation context.  In particular, the
rebuild rules capture binder frames.  If only binder frames remain, then
execution is complete.  When either of the other frames is found, then a
reduction is performed.  On the other hand, the need rule captures the
evaluation context up to the binder that matches the variable whose value is
needed.

These actions of the abstract machine can be recast in the language of
delimited control capture.
First, the need rule uses the identity of its variable, which must be an active
variable, to delimit the context it captures. The well-formedness conditions
from Section~\ref{sec:hygiene} guarantee that each binder frame binds a unique
variable, so each active variable acts as a unique delimiter.
Second, the rebuild rule uses the nearest non-binder frame to delimit the
context it captures.  This means that rebuild operates as though the operand
frames, the cont frames, and the top of the evaluation context share a common
delimiter.  This guarantees that only binder frames are captured (as is
stipulated in the rules).

In short, call-by-need evaluation captures partial evaluation contexts.  These
partial evaluation contexts correspond to delimited continuations, and there
are two different kinds of delimitation, redex-based (for rebuild) and 
binder-based (for need). 

It is useful to also consider how the machine manipulates these delimited
continuations.  Each reduce rule in Figure~\ref{fig:transitions} immediately
pushes the context associated with an answer onto the current evaluation
context.  In this manner, binders are consistently moved above the point of
evaluation.  The reduce rule then operates on the value part of the answer and
the associated cont (for $D.1$) or term (for $D.2$).

Although each reduce rule pushes binders onto the evaluation context, only
the $D.2$ rule creates new binders.  The variable bound by the answer's
underlying lambda abstraction may already be a member of the set $X$, in which
case it must be alpha-converted to a fresh name with respect to the set
$X$.  Also note that if $\lambda x.t$ is alpha converted to $\lambda
x'.t[x'/x]$, the body under call-by-value satisfies the equation $t[x'/x] =
(\lambda x.t)\:x'$.  Since we are using the identifiers $x'$ as delimiters,
and we never turn the binder frame $[(\lambda x'.[])\;t]$ back into a term, we
can replace fresh variables $x'$ with fresh \emph{prompts}~\cite{balat04sums}.

\begin{figure}
  \centering\small
  \begin{boxedarray}{@{}l@{}}
    \text{Let $s$ be a distinguished identifier:} \\ \\

      \mathcal{N}^P|[ t |]  =  \<runCC> 
      (\<let> s = \<newPrompt>\; \<in> \<pushPrompt> s \; \mathcal{N}|[ t |])
           \\ \\

      \mathcal{N}|[ x |]  = \mathsf{need}\;x \\ \\

      \mathcal{N}|[ t_1\;t_2 |] =  
      \begin{array}[t]{@{}l@{}}
        \mathsf{do}\;v_a <= \mathcal{N}|[t_1|] \\
        \mathsf{in}\; 
        \begin{array}[t]{@{}l@{}}
          \<let> x_p = \<newPrompt> \\
          \<in> 
          \begin{array}[t]{@{}l@{}}
            \mathsf{delay}\;\mathcal{N}|[t_2|]\;\mathsf{as}\;x_p \;
            \mathsf{in}\;(v_a \;x_p)
          \end{array} 
        \end{array} 
      \end{array} 
      \\ \\

      \mathcal{N}|[ \lambda x.t |]  = 
      \mathsf{return}\; \lambda x. \mathcal{N}|[ t |] \\ \\

      \hline
      \\

      \mathsf{return}\;v_a \equiv     
      \<withSubCont> s \; \lambda k_a.\Braket{k_a,v_a}  \\ \\

      \mathsf{do}\;x <= t_1 \;\mathsf{in}\; t_2 \equiv
      \begin{array}[t]{@{}l@{}}
          \<let> \Braket{k_a,x} = \<pushPrompt> s \; t_1 \\
          \!\!\<in> \<pushSubCont> k_a \; t_2
      \end{array} \\ \\

      \mathsf{delay}\; t_1\; \mathsf{as}\; x\;\mathsf{in}\; t_2 \equiv
      \begin{array}[t]{@{}l@{}}
            \<let> f_k = \<pushPrompt> x\; t_2\\
            \!\!\<in> f_k\; \lambda (). t_1 
      \end{array} \\ \\
 
      \mathsf{force}\;f \equiv f\;() \\ \\

      \mathsf{need}\;x \equiv\;
      \begin{array}[t]{@{}l@{}}
          \<withSubCont> x \; \lambda k.\\
          \quad\lambda f_{th}. 
      \begin{array}[t]{@{}l@{}}

          \mathsf{do}\;v_a <= \mathsf{force}\;f_{th} \\
          \mathsf{in}\;
          \begin{array}[t]{@{}l@{}}
            \mathsf{delay}\;(\mathsf{return}\;v_a)\;\mathsf{as}\;x \\
            \mathsf{in}\;\<pushSubCont> k \; (\mathsf{return}\;v_a) \\ 
          \end{array} 
        \end{array} 
      \end{array} 
      \\ \\

  \end{boxedarray}
  
   \caption{Translating CBN to CBV+Control}
  \label{fig:cbn-simulation}
\end{figure}

From these observations, we construct the simulation in
Figure~\ref{fig:cbn-simulation}.
The simulation can be understood as a direct encoding of the abstract machine
semantics for call-by-need.  To execute a program, $\mathcal{N}^P|[t|]$, the
transformation uses $\<runCC>\!$ to initiate a control-based
computation, acquires a fresh prompt, and binds it to a distinguished variable
$s$.  This prompt is the \emph{redex prompt}, which is used to delimit every
continuation that denotes a redex.

To expose the conceptual structure of the simulation, we define five syntactic
macros, $\mathsf{do}$, $\mathsf{return}$, $\mathsf{delay}$, $\mathsf{force}$,
and $\mathsf{need}$.  We accord no formal properties to them: they merely
simplify the presentation.  The $\mathsf{return}$ macro captures the nearest
subcontinuation that is delimited by the redex prompt $s$. Since the $s$
delimiter appears before every reduction, the captured continuation is
guaranteed to contain only code equivalent to binder frames.  The translation
returns a tuple containing the subcontinuation and the argument to
$\mathsf{return}$, which must be a value; the tuple represents an answer.  So
the translation rule for lambda abstractions, $\mathcal{N}|[\lambda x.t|]$,
literally simulates the rebuild rules.

The $\mathsf{do}$ macro executes a term $t_1$ under the current continuation
extended with the redex prompt. If the term returns an answer $\Braket{k_a,x}$
it immediately pushes the subcontinuation part and continues execution, binding
the value part to the variable $x$.  As such, the translation rule for
applications, $\mathcal{N}|[t_1\:t_2|]$, executes $|[t_1|]$ and binds the
resulting operator to $v_a$. The answer binders are pushed by the $\mathsf{do}$
macro, which starts the simulation of the $D.2$ rule.

The remainder of the $D.2$ rule is subtle. In the abstract machine, binder
frame variables delimit the need rules.  Since the delimited continuation
framework relies on prompts to delimit continuations, fresh prompts literally
substitute for variables~\cite{kiselyov06dynamic}.  The translation uses
$\<newPrompt>$ to acquire a fresh prompt $x_p$ and then uses the
$\mathsf{delay}$ macro to simulate pushing a binder frame: the context
$\mathsf{delay}\;t \;\mathsf{as}\;x\;\mathsf{in}\;[]$ is analogous to the
binder frame $[(\lambda x.[])\;t]$.  The $\mathsf{delay}$ macro anticipates
that its body returns a function $f_k$ that expects the delayed argument, so it
applies $f_k$ to a suspension of $t$.  As we see below, the function $f_k$ is a
cont $(\kappa x.E)$.

In the context of $\mathsf{delay}$, the simulation executes $v_a\:x_p$.  Since
alpha conversion of $\lambda x.t$ can be written $(\lambda x_p.t[x_p/x])$, the
term~$v_a\:x_p$ is analogous to $(\lambda x.t)\:x_p = t[x_p/x]$: it substitutes
a fresh prompt for a fresh variable.

The $\mathsf{need}$ macro, which defines the translation rule for variables,
$\mathcal{N}|[x|]$, captures the continuation delimited by $x$ (which had
better be a prompt!) and returns a function $\lambda f_{th}.\dots$ that closes
over both $x$ and the captured continuation $k$. This function is the cont
$\kappa x.E$, with $x$ modeling the bound variable of the same name, and
continuation $k$ modeling $E$. The function expects the binder frame $[(\lambda
x.[])\;t]$, which is now at the top of the current continuation, to pass it the
suspension $\lambda ().\mathcal{N}|[t|]$.  The simulation forces the
suspension, and the $\mathsf{do}$ macro pushes the resulting answer binders and
binds $v_a$ to the underlying value.  Pushing the answer binders begins the
simulation of the $D.1$ rule.

The simulation of $D.1$ delays a computation that immediately returns the
result $v_a$ of evaluating the term $t$, pushes the continuation $k$ associated
with the cont, and returns $v_a$ to the extended continuation.  Now any
subsequent evaluation of $x$ immediately returns the memoized value $v_a$
instead of recomputing $t$.  This yields an answer $\Braket{k_a,v_a}$ where
$k_a$ is an empty subcontinuation.  The value $v_a$ is delayed exactly as
before and is also returned from the properly extended continuation.  This part
of the translation bears close resemblance to the paradoxical $\mathsf{Y}$
combinator~\cite{curry-n-feys}, suggesting that the simulation requires
recursive types~\cite{shan07static}.

\section{Correctness of the Simulation}

We prove correctness of the simulation relative to the machine semantics.
Since we already proved correctness of the machine semantics relative to
standard-order reduction, the result is a proof that our simulation
provides a continuation semantics for call-by-need.

The previous discussion provides an informal justification for the structure of
the call-by-need simulation.  To prove the correctness of the simulation, we
appeal to the continuation semantics for delimited
control~\cite{dybvig07monadic}.  This semantics is completely standard for the
terms of the lambda calculus.  Figure~\ref{fig:dc-combinators} presents the
interesting parts of the semantics.  All CPS terms take a standard continuation
$\kappa$, but the control combinators also take a \emph{metacontinuation}
$\gamma$, which is a list of continuations and prompts, and a global prompt
counter $q$.  The base continuation $\kappa_0$ delimits each proper
continuation and directs evaluation up the metacontinuation, discarding any
intervening prompts. Given a CPS program $t$, the expression
$t\;\kappa_0\;[\;]\;0$ runs it.

\begin{figure}[t]
  \centering\small
  \begin{boxedarray}{@{}l@{}}
    newPrompt_c = \lambda\kappa.\lambda\gamma.\lambda q.\kappa\;q\;\gamma\;(q+1)
    \\
    withSubCont_c =  \lambda p.\lambda f.\lambda\kappa.\lambda\gamma.
    f\;(\kappa:\gamma^p_\uparrow)\;\kappa_0\;\gamma^p_\downarrow \\
    pushPrompt_c = \lambda p.\lambda t.\lambda\kappa.\lambda\gamma.
    t\;\kappa_0\; (p:\kappa:\gamma)  \\
    pushSubCont_c = \lambda\gamma'.\lambda t.\lambda\kappa.\lambda\gamma.
    t\;\kappa_0\; (\gamma'\! ++ (\kappa:\gamma)) \\
    \kappa_0 = \lambda v.\lambda\gamma\lambda q.\mathcal{K}(v,\gamma,q) \\
    \\
    \mathcal{K}(v,[\;],q) = v\\
    \mathcal{K}(v,p:\gamma,q) = \mathcal{K}(v,\gamma,q) \\
    \mathcal{K}(v,\kappa:\gamma,q) = \kappa\;v\;\gamma\;q \\
    \end{boxedarray}
   \caption{Delimited Control Combinators}
  \label{fig:dc-combinators}
\end{figure}

To prove correctness, we compose $\mathcal{N}|[\cdot|]$ with the delimited
continuation semantics to produce a translation $\Lambda |[\cdot|]$ to the
$\lambda_{\beta\eta}$ calculus augmented with arithmetic, lists, and the
operator $\mathcal{K}$ defined in Figure~\ref{fig:dc-combinators}.  We also
give each abstract machine configuration a denotation, defined in terms of
name-indexed denotations for its constituents $\mathcal{D}|[\cdot|]_X$ (see
Figures~\ref{fig:cbn-denote-term} through \ref{fig:cbn-denote-redex}).

\begin{figure}[ht]
  \centering\small
  \begin{boxedarray}{@{}l@{}}

    \mathcal{D}|[ t |]_X  =
    \Lambda |[ t |]\overline{[\iota(x_i,X)/x_i]}  \\ \\

      \Lambda^P |[ t|] =
      \Lambda |[t|]\; \kappa_0 \; (0:[\;])\; 1 \\ \\

      \Lambda |[ x |]  =  \\
      \begin{array}[t]{@{}l@{}}
\mathit{withSubCont_c}\;x \\
\quad \lambda k_x.\lambda k_1.k_1  \\
\quad\quad\: \lambda f_{th}.\lambda k_2. \\
\qquad\quad\;\: \mathit{pushPrompt_c}\; 0\; (f_{th}\;()) \\
\qquad\qquad\;(\lambda \Braket{k_a,v_a}\!.\\
\qquad\qquad\quad \mathit{pushSubCont_c}\; k_a \\
\qquad\qquad\qquad(\lambda k_3.\\
\qquad\qquad\qquad\;\:\mathit{pushPrompt_c}\; x \\
\qquad\qquad\qquad\quad(\mathit{pushSubCont_c}\; k_x \\
\qquad\qquad\qquad\qquad(\mathit{withSubCont_c}\; 0\; 
\lambda k_a. \lambda k. k \Braket{k_a, v_a})) \\
\qquad\qquad\qquad\quad(\lambda f_k.f_k\;
 (\lambda ().\mathit{withSubCont_c}\; 0\; \\
\qquad\qquad\qquad\qquad\qquad\qquad\quad\;
 \lambda k_a.\lambda k. k \Braket{k_a, v_a}) \\
\qquad\qquad\qquad\quad\qquad\qquad\!\!k_3)) \\
\qquad\qquad\qquad k_2) \\
      \end{array}  \\ \\

      \Lambda |[ t_1\;t_2 |] =  
      \begin{array}[t]{@{}l@{}}
\lambda k_1.\mathit{pushPrompt_c}\; 0\; \Lambda |[ t_1 |] \\
\qquad(\lambda\Braket{k_a,v_a}\!. \\
\qquad\quad\mathit{pushSubCont_c}\; k_a \\
\qquad\qquad(\lambda k_2. \\
\qquad\qquad\quad\mathit{newPrompt_c} \\
\qquad\qquad\qquad\lambda x_p.\mathit{pushPrompt_c}\; x_p\; (v_a\; x_p)  \\
\qquad\qquad\qquad\qquad\quad
(\lambda f_k.f_k\; (\lambda (). \Lambda |[ t_2 |])\; k_2)) \\
\qquad\qquad k_1) \\
      \end{array} \\ \\

      \Lambda |[ \lambda x.t |]  = 
      \begin{array}[t]{@{}l@{}}
\mathit{withSubCont_c}\;0 \;\lambda k_a.\lambda k. k
 \braket{k_a, \lambda x. \Lambda \Lbrack t \Rbrack} \\ 
      \end{array} 
  \end{boxedarray}
  
   \caption{Denotations for Terms}
  \label{fig:cbn-denote-term}
\end{figure}

Denotations of machine configurations are constructed from their components:
the configuration's focus $?$, context $E$, and list of names $X$.  A machine
configuration denotes the translation of its focus applied to three arguments:
the base continuation $\kappa_0$ as its starting continuation, the denotation
of its context, bounded by the redex delimiter $0$, as the metacontinuation,
and the size $\lvert X \rvert$ of $X$ plus 1 as its initial prompt.  The redex
delimiter attached to the metacontinuation handles the case when an answer
subsumes the entire context by returning the answer as the result.  The
denotation of the terminal machine configuration $\Braket{X|\answer{E,v}}$ is
treated separately to show how it corresponds directly to a final answer.

\begin{figure}[t]
  \centering\small
  \begin{boxedarray}{@{}l@{}}
      \iota (x_i,X) = \iota(x_i,[x_1,x_2,\dots,x_i,\dots,x_n])  =  i  \\ \\
      \lvert X \rvert = \lvert [x_1,x_2,\dots,x_i,\dots,x_n] \rvert  =  n  \\ \\
      \Lambda |[\Braket{X|E,?}_c |]  = 
      \mathcal{D}|[ ? |]_X \; \kappa_0 \; (\mathcal{D}|[ E |]_X ++ (0:[\;])) \;
      (\lvert X \rvert + 1)  \quad c\in\set{d,f,b,n} \\ \\
      \Lambda |[\Braket{X|\answer{E,\lambda x.t}}|] = 
      \braket{\mathcal{D}|[ E |]_X,\lambda x.\mathcal{D}|[ t |]_X}

    \end{boxedarray}
   \caption{Denotations for Names and Configurations}
  \label{fig:cbn-denote-misc}
\end{figure}

Our semantic translation takes advantage of $X$ being a proper list of unique
names.  Free active variables denote prompts in our translation, and since $0$
is the redex delimiter, we assign to each variable its $1$-based index in $X$.
We use $\lvert X \rvert + 1$ as the global prompt counter to ensure that no
future prompts conflict with the current active variable denotations, thereby
guaranteeing hygiene (see Section~\ref{sec:hygiene}).

Each evaluation context frame denotes a two-element metacontinuation consisting
of a prompt and a proper continuation.  The prompt for a binder frame is the
prompt translation $\iota(x,X)$ of the bound variable $x$.  The cont and
operand frames have redex prompts~$0$.  These prompts guarantee that answer
building operations will arrive at the innermost redex.  Each continuation
function specializes a subexpression of the CPS translation for terms
$\mathcal{D}|[\cdot|]_X$ with the denotations of the context frame's parts.  Compare, for
instance, the denotation of an application, $t_1\;t_2$, to that of an operand
frame, $[[]\;t_2]$.  The application term pushes the global prompt, and
executes $t_1$ in the context of a continuation that receives an answer
$\Braket{k_a,v_a}$.  The denotation of the operand frame is a metacontinuation
containing the same prompt and continuation.

\begin{figure}[ht]
  \centering\small
  \begin{boxedarray}{@{}l@{}}

    \mathcal{D}|[ E \comp [f] |]_X  =
    \mathcal{D}|[[f]|]_X ++ \mathcal{D}|[E|]_X  \\ \\

    \mathcal{D}|[ [\;] |]_X  =  [\;]  \\ \\

    \mathcal{D}|[ [\#] |]_X  =  \kappa_0:[\;]  \\ \\

    \mathcal{D}|[ [[]\; t_2] |]_X  =   0 : k' : [\;]  \\
        \text{where } k' = 
      \begin{array}[t]{@{}l@{}}
\lambda\Braket{k_a,v_a}\!. \\
\quad\mathit{pushSubCont_c}\; k_a \\
\qquad(\lambda k_2. \\
\qquad\quad\mathit{newPrompt_c} \\
\qquad\qquad\lambda x_p.\mathit{pushPrompt_c}\; x_p\; (v_a\; x_p)  \\
\qquad\qquad\qquad\quad(\lambda f_k.f_k\; (\lambda (). \mathcal{D}|[ t_2 |]_X)\; k_2)) \\
\qquad \kappa_0
      \end{array}  \\ \\ 

      \mathcal{D}|[ [(\lambda x.[])\; t_2] |]_X  =  \iota(x,X) : k' : [\;]  \\
        \text{where } k' = 
      \lambda f_k.f_k\; (\lambda (). \mathcal{D}|[ t_2 |]_X)\; \kappa_0 \\ \\

      \mathcal{D}|[ [(\kappa x.E)\;[]] |]_X  =   0 : k' : [\;]  \\
        \text{where } k' = \\
      \begin{array}[t]{@{}l@{}}
\;\lambda \Braket{k_a,v_a}\!.\\
\quad \mathit{pushSubCont_c}\; k_a \\
\qquad(\lambda k_3.\\
\qquad\;\:\mathit{pushPrompt_c}\; \iota(x,X) \\
\qquad\quad(\mathit{pushSubCont_c}\; \mathcal{D}|[E|]_X \\
\qquad\qquad(\mathit{withSubCont_c}\; 0\; 
\lambda k_a. \lambda k. k \Braket{k_a, v_a})) \\
\qquad\quad(\lambda f_k.f_k\;
 (\lambda ().\mathit{withSubCont_c}\; 0\;
\lambda k_a.\lambda k. k \Braket{k_a, v_a}) \\
\qquad\quad\qquad\qquad\!\!k_3)) \\
\qquad \kappa_0
      \end{array}

  \end{boxedarray}
  
   \caption{Denotations for Evaluation Contexts}
  \label{fig:cbn-denote-context}
\end{figure}

A redex denotes a CPS'ed term that closes over the denotations of its
constituents and implements the corresponding reduction step.

\begin{figure}
  \centering\small
  \begin{boxedarray}{@{}l@{}}
      \mathcal{D}|[ (\kappa x_1.E_1)\;\answer{E_2,\lambda x_2.t} |]_{X}  = \\
      \begin{array}[t]{@{}l@{}}
 \mathit{pushSubCont_c}\; \mathcal{D}|[ E_2 |]_X \\
\quad(\lambda k_3.\\
\quad\;\:\mathit{pushPrompt_c}\; \iota(x_1,X) \\
\quad\quad(\mathit{pushSubCont_c}\; \mathcal{D}|[ E_1 |]_X \\
\quad\qquad(\mathit{withSubCont_c}\; 0\; 
\lambda k_a. \lambda k. k \Braket{k_a, \lambda x_2.\Lbrack t \Rbrack_X})) \\
\qquad(\lambda f_k.f_k\;
 (\lambda ().\mathit{withSubCont_c}\; 0\;
\lambda k_a.\lambda k. k \Braket{k_a, \lambda x_2.\Lbrack t \Rbrack_X}) \\
\qquad\qquad\qquad\!\!k_3)) \\
      \end{array}  \\ \\

      \mathcal{D}|[ \answer{E,\lambda x.t_1}\;t_2 |]_X  =  \\
      \begin{array}[t]{@{}l@{}}
\mathit{pushSubCont_c}\; \mathcal{D}|[ E |]_X \\
\quad(\lambda k_2. \\
\quad\quad\mathit{newPrompt_c} \\
\quad\qquad\lambda x_p.\mathit{pushPrompt_c}\; x_p\;
((\lambda x.\mathcal{D}|[ t_1 |]_X)\; x_p)  \\
\quad\qquad\qquad\quad
(\lambda f_k.f_k\; (\lambda (). \mathcal{D}|[ t_2 |]_X)\; k_2)) \\
      \end{array} 
  \end{boxedarray}
  
   \caption{Denotations for Redexes}
  \label{fig:cbn-denote-redex}
\end{figure}

\topic{small modifications to the machine semantics to facilitate proving}
To facilitate our proof of correctness, we make a slight change to the machine
semantics.  In the machine, composing an empty context with the current
context is an identity operation.  The continuation semantics do not share
this property.  During execution, an empty continuation is denoted by the base
continuation $\kappa_0$.  If a continuation is captured or pushed in the
context of an empty continuation, then the empty continuation will be captured
as part of the metacontinuation or pushed onto the current metacontinuation
before reinstating the pushed continuation.  In short, the call-by-need machine
semantics guarantees that $E \comp [\;] = E$, but the continuation semantics do
not prove that $\kappa_0:\gamma = \gamma$.  Dybvig et al. discuss the notion of
proper tail recursion for delimited continuations.  Their operational
characterization of proper tail recursion corresponds to the latter equation.

To remove this mismatch, we add a \emph{ghost} frame $[\#]$ to our definition
of evaluation contexts.  The ghost frame denotes the metacontinuation
$\kappa_0:[\;]$.  We also extend the unplug operation on evaluation contexts
such that it discards ghost frames: $\mtoc |[E \comp [\#]|] =~\mtoc|[E|]$.
Finally, we alter the right hand side of transition rules that grab and push
continuations to pair ghost frames with composed evaluation contexts in a
manner consistent with the continuation semantics. For instance, the updated
$D.1$ rule is as follows\footnote{The exact placement of ghost frames falls
  right out of the correctness proof.}:
\begin{equation*}
\tag{D.1}  \reduce{X | E_1, (\kappa x.E_2)\;\answer{E_3,v}} \nam\ 
  \rebuild{X | E_1 \comp [\#] \comp E_3 \comp [(\lambda x.[])\; v] 
    \comp [\#] \comp E_2, v}
\end{equation*}
These modifications do not alter the observable behavior of the machine while
modeling the property that pushing empty frames has meaning in the
continuation semantics.

Given these denotations, it is straightforward to prove correctness of the
simulation relative to the abstract machine.

\begin{thm}
  If $t$ is a closed term, then
  $\Lambda^P |[t|] = 
\Lambda |[\refocus{\emptyset | [\;],t} |]$.
\end{thm}
\begin{proof}
$ \Lambda^P |[t|] = \mathcal{D}|[ t |]_\emptyset \; \kappa_0 \; (0:[\;])\; 1  
  = \Lambda |[\refocus{\emptyset | [\;],t} |].$
\end{proof}

\begin{thm}
  If $C_1 \nam\ C_2$ then $\Lambda |[C_1|] = \Lambda |[C_2|]$.
\end{thm}
\begin{proof}
  By cases on $\nam\ $.  The proof utilizes beta, eta and $\mathcal{K}$
  equivalences to establish correspondences.
\end{proof}

\section{Simulating Extensions}

Many straightforward language extensions also have straightforward simulations
under the above model.

Simulating $\<let>\!\!$ bindings essentially performs the same operations as
immediately applying a lambda abstraction.  
\begin{displaymath}
\mathcal{N}|[\<let> x = t_1 \<in>\; t_2|] =   
        \begin{array}[t]{@{}l@{}}
          \<let> x = \<newPrompt> \\
          \!\!\<in>\; 
          \begin{array}[t]{@{}l@{}}
            \mathsf{delay}\;\mathcal{N}|[t_1|]\;\mathsf{as}\;x \;\\
            \mathsf{in}\;\mathcal{N}|[t_2|]
          \end{array} 
        \end{array} 
\end{displaymath}
In contrast to the application rule, the variable $x$ is directly assigned a
fresh prompt, rather than binding it to an auxiliary variable $x_p$.  The body
of the let can be interpreted in place and substitution of the prompt is
implicit since $x$ is already free in $t_2$.

The translation for basic constants is analogous to that for lambda
abstractions:  the constant must be returned to the next redex.
\begin{displaymath}
  \mathcal{N}|[ c |] = \mathsf{return}\;c
\end{displaymath}
For this translation, we assume that the call-by-value language provides
the same constants as the call-by-need language.

The translation for function constants is as follows:
\begin{displaymath}
\mathcal{N}|[ f\;t |] = 
      \begin{array}[t]{@{}l@{}}
        \mathsf{do}\;v_a <= \mathcal{N}|[t|] \\
        \mathsf{in}\;\mathsf{return}\;(f\;v_a)
      \end{array}   
\end{displaymath}
Interpreting a function constant application forces its argument and then
acts on the value that is ultimately produced.  Since function expressions
yield values, the result is immediately returned.

The translation for $\mathtt{cons}$ acquires two fresh prompts, uses them to
delay the argument to $\mathtt{cons}$, and stores them as a pair.
\begin{displaymath}
\mathcal{N}|[ \mathtt{cons}\;t_1\;t_2 |] = 
      \begin{array}[t]{@{}l@{}}
        \<let> x_1 = \<newPrompt> \<in> \\
        \<let> x_2 = \<newPrompt> \<in> \\
        \mathsf{delay}\; \mathcal{N}|[t_1|] \;\mathsf{as}\; x_1 \;\mathsf{in} \\
        \mathsf{delay}\; \mathcal{N}|[t_2|] \;\mathsf{as}\; x_2 \;\mathsf{in} \\
        \mathtt{cons}\;x_1\;x_2
      \end{array}   
\end{displaymath}

The translations for $\mathtt{car}$ and $\mathtt{cdr}$ evaluate their
argument, retrieve a prompt from the resulting pair, and demand its value.
\begin{displaymath}
\begin{array}[t]{@{}l@{}}
\mathcal{N}|[ \mathtt{car}\;t |] = 
      \begin{array}[t]{@{}l@{}}
        \mathsf{do}\;v_p <= \mathcal{N}|[t|] \\
        \mathsf{in}\;\mathsf{need}\;(\mathtt{car}\;v_p)
      \end{array}  \\ \\

\mathcal{N}|[ \mathtt{cdr}\;t |] = 
      \begin{array}[t]{@{}l@{}}
        \mathsf{do}\;v_p <= \mathcal{N}|[t|] \\
        \mathsf{in}\;\mathsf{need}\;(\mathtt{cdr}\;v_p)
      \end{array}  \\ \\
\end{array}
\end{displaymath}

\section{Conclusions}

In this paper, we expose and examine the operational structure of lazy
evaluation as embodied in call-by-need semantics.  We present this
understanding in two ways: as an abstract machine whose operational behavior
involves control capture, and as a simulation of call-by-need under
call-by-value plus delimited control operations. Delimited control can be
used to simulate a global heap, but our particular simulation uses delimited
control operations to manage laziness locally, just like the calculus reduction
rules.

The artifacts of this investigation provide new tools for increasing our
understanding of lazy evaluation and its connections to control.  The abstract
machine could be used to establish connections to heap-based implementations of
call-by-need, and possibly modern graph-reduction based
formulations~\cite{peytonjones89spineless}.  In fact it seems that the calculus
and abstract machine may point out new structural and dynamic invariants that
are inherent to call-by-need evaluation but are hidden in the unstructured
representations of heaps.

The abstract machine and simulation might also provide new opportunities for
reasoning about the correctness of transformations applied to call-by-need
programs.  Surely the calculus provides the same equational reasoning powers as
the abstract machine.  However the machine may enable researchers to more
easily conceive transformations and justifications that are not as easy to
recognize in the reduction semantics.  Our simulation might be connected to
that of~\citet{OKASAKI94}.  The simulation might suggest new mechanisms by
which to embed call-by-need evaluation within call-by-value programs.

One significant difference between the two formulations of call-by-need lambda
calculi~\cite{maraist98need,ariola97need} is the status of variables.  Maraist
et al. consider variables to be values, whereas Ariola and Felleisen do not.
Ultimately, \citet{maraist98need} prove standardization against a
standard-order relation that does not consider variables to be values.  This
paper sheds no light on the inclusion of variables among the values, however it
demonstrates in stark detail the consequences of the latter design.  In the
abstract machine, the transition rules for lambda terms, namely the rebuild
rules, differ significantly from the transition rules for variables, the need
rules.  A similar distinction can be seen simply by observing the complexity of
their respective translations.  In short, our semantics interpret variables as
memoized computations rather than values.  Variables can be treated as 
values under deterministic call-by-value and call-by-name reduction; it remains
an open question whether the same could be achieved for call-by-need and if so
what its operational implications would be.

Our results reveal that a proliferation of semantic frameworks---reduction
semantics, machine semantics, etc---is a boon and not a crisis.  The reduction
semantics of call-by-need elegantly and mysteriously encode a rich semantics
whose broad implications can be seen in equivalent machine semantics and
continuation semantics.  As such, our work provides new perspectives from which
to reason about call-by-need, delimited control, and their respective
expressive powers.

\section{Acknowledgements}
We thank Daniel P. Friedman, Roshan James, William Byrd, Michael Adams, and the
rest of the Indiana University Programming Languages Group, as well as Jeremy
Siek, Zena Ariola, Phil Wadler, Olivier Danvy, and anonymous referees for
helpful discussions and feedback on this work.

\bibliographystyle{acmtrans}
\bibliography{need}

\end{document}

%% file: definitions.tex
\usepackage[leqno]{amsmath}
\usepackage{amssymb}
\usepackage{amsthm}
\usepackage{braket}
\usepackage{bbold}
\usepackage{semantic}
\usepackage{mathpartir}
\usepackage{stmaryrd}
\usepackage{mathbbol} 

\newcommand{\union}[0]{\ensuremath{\cup}}

\newcommand{\desc}[1]{\ensuremath{\text{(#1)}}}

\newcommand{\seq}[1]{\ensuremath{\overline{#1}}}

\newcommand{\produce}{\ensuremath{::=}}
\newcommand{\hole}{\square}

\newenvironment{ltransitionrule}
  {\begin{array}{lrcll}}
  {\end{array}}

\newsavebox{\saveboxedarray}
\newenvironment{boxedarray}[1]
  {\begin{lrbox}{\saveboxedarray}\begin{math}\begin{array}{#1}}
  {\end{array}\end{math}\end{lrbox}\fbox{\usebox{\saveboxedarray}}}

\newenvironment{inferbox*}[0]
  {\begin{minipage}{\columnwidth}\begin{mathpar}}
  {\end{mathpar}\end{minipage}}

\newcommand{\mtoc}[0]{\mathcal{C}}

\theoremstyle{plain}

\newtheorem{theorem}{Theorem}

\newtheorem{corollary}[theorem]{Corollary}

\theoremstyle{remark}
 \newtheorem*{case}{Case}

\mathlig{[]}{\hole}
\mathlig{[}{\lbrack}
\mathlig{]}{\rbrack}
\mathlig{|}{\mid}

\mathlig{++}{\texttt{+\kern-0.03em+}}

\newcommand{\mapsTo}[0]{%
  \mbox{\ensuremath{\makebox[0pt][l]{\hspace{7pt}$\rightarrow$}\longmapsto}}}

\newcommand{\nam}[0]{\ensuremath{\longmapsto_{nam}}}
\newcommand{\Nam}[0]{\ensuremath{\mapsTo_{nam}}}

\newcommand{\sr}[0]{\ensuremath{\longmapsto_{sr}}}
\newcommand{\Sr}[0]{\ensuremath{\mapsTo_{sr}}}

\renewcommand{\comp}[0]{\ensuremath{\circ}}

\newcommand{\answer}[1]{\llfloor#1\rrfloor}
\newcommand{\refocus}[1]{\Braket{#1}_f}
\newcommand{\rebuild}[1]{\Braket{#1}_b}
\newcommand{\reduce}[1]{\Braket{#1}_d}
\newcommand{\need}[1]{\Braket{#1}_n}

\newcommand{\notion}[0]{\ensuremath{\rightarrow_{\mathrm{need}}}}

\reservestyle{\command}{\textsl}
\reservestyle{\bfcommand}{\textbf}
\bfcommand{wf[\;wf],ok[\;ok]}


%% file: need.bbl
\begin{thebibliography}{}

\bibitem[\protect\citeauthoryear{Abadi, Cardelli, Curien, and L{\'e}vy}{Abadi
  et~al\mbox{.}}{1991}]{abadi91explicit}
{\sc Abadi, M.}, {\sc Cardelli, L.}, {\sc Curien, P.-L.}, {\sc and} {\sc
  L{\'e}vy, J.-J.} 1991.
\newblock Explicit substitutions.
\newblock {\em Journal of Functional Programming\/}~{\em 1,\/}~4, 375--416.

\bibitem[\protect\citeauthoryear{Ariola and Klop}{Ariola and
  Klop}{1994}]{ariola94rewriting}
{\sc Ariola, Z.} {\sc and} {\sc Klop, J.~W.} 1994.
\newblock Cyclic lambda graph rewriting.
\newblock In {\em Logic in Computer Science ({LICS} '94)}. IEEE Computer
  Society Press, Los Alamitos, Ca., USA, 416--425.

\bibitem[\protect\citeauthoryear{Ariola and Felleisen}{Ariola and
  Felleisen}{1997}]{ariola97need}
{\sc Ariola, Z.~M.} {\sc and} {\sc Felleisen, M.} 1997.
\newblock The call-by-need lambda calculus.
\newblock {\em Journal of Functional Programming\/}~{\em 7,\/}~3 (May),
  265--301.

\bibitem[\protect\citeauthoryear{Ariola and Klop}{Ariola and
  Klop}{1997}]{ariola97recursion}
{\sc Ariola, Z.~M.} {\sc and} {\sc Klop, J.~W.} 1997.
\newblock Lambda calculus with explicit recursion.
\newblock {\em Information and Computation\/}~{\em 139,\/}~2, 154--233.

\bibitem[\protect\citeauthoryear{Ariola, Maraist, Odersky, Felleisen, and
  Wadler}{Ariola et~al\mbox{.}}{1995}]{ariola95need}
{\sc Ariola, Z.~M.}, {\sc Maraist, J.}, {\sc Odersky, M.}, {\sc Felleisen, M.},
  {\sc and} {\sc Wadler, P.} 1995.
\newblock A call-by-need lambda calculus.
\newblock In {\em POPL '95: Proceedings of the 22nd ACM SIGPLAN-SIGACT
  Symposium on Principles of Programming Languages}. ACM Press, New York, NY,
  USA, 233--246.

\bibitem[\protect\citeauthoryear{Balat, Di~Cosmo, and Fiore}{Balat
  et~al\mbox{.}}{2004}]{balat04sums}
{\sc Balat, V.}, {\sc Di~Cosmo, R.}, {\sc and} {\sc Fiore, M.} 2004.
\newblock Extensional normalisation and type-directed partial evaluation for
  typed lambda calculus with sums.
\newblock In {\em {POPL} '04: Proceedings of the 31st {ACM} {SIGPLAN-SIGACT}
  Symposium on Principles of Programming Languages}. ACM, New York, NY, USA,
  64--76.

\bibitem[\protect\citeauthoryear{Barendregt}{Barendregt}{1981}]{barendregt}
{\sc Barendregt, H.~P.} 1981.
\newblock {\em The Lambda Calculus, its Syntax and Semantics}.
\newblock North-Holland, Amsterdam, NL.
\newblock Studies in Logic and the Foundations of Mathematics.

\bibitem[\protect\citeauthoryear{Biernacka and Danvy}{Biernacka and
  Danvy}{2007}]{biernacka07framework}
{\sc Biernacka, M.} {\sc and} {\sc Danvy, O.} 2007.
\newblock A concrete framework for environment machines.
\newblock {\em ACM Transactions on Computational Logic\/}~{\em 9,\/}~1, 6.

\bibitem[\protect\citeauthoryear{Biernacki, Danvy, and chieh Shan}{Biernacki
  et~al\mbox{.}}{2005}]{Biernacki20057}
{\sc Biernacki, D.}, {\sc Danvy, O.}, {\sc and} {\sc chieh Shan, C.} 2005.
\newblock On the dynamic extent of delimited continuations.
\newblock {\em Information Processing Letters\/}~{\em 96,\/}~1, 7 -- 17.

\bibitem[\protect\citeauthoryear{Curry and Feys}{Curry and
  Feys}{1958}]{curry-n-feys}
{\sc Curry, H.~B.} {\sc and} {\sc Feys, R.} 1958.
\newblock {\em Combinatory Logic, Volume {I}}.
\newblock Studies in Logic and the Foundations of Mathematics. North-Holland,
  Amsterdam.
\newblock Second printing 1968.

\bibitem[\protect\citeauthoryear{Danvy and Filinski}{Danvy and
  Filinski}{1990}]{danvy90abstracting}
{\sc Danvy, O.} {\sc and} {\sc Filinski, A.} 1990.
\newblock Abstracting control.
\newblock In {\em LFP '90: Proceedings of the 1990 ACM Conference on LISP and
  Functional Programming}. ACM, New York, NY, USA, 151--160.

\bibitem[\protect\citeauthoryear{Danvy, Millikin, Munk, and Zerny}{Danvy
  et~al\mbox{.}}{2010}]{danvy10need}
{\sc Danvy, O.}, {\sc Millikin, K.}, {\sc Munk, J.}, {\sc and} {\sc Zerny, I.}
  2010.
\newblock Defunctionalized interpreters for call-by-need evaluation.
\newblock In {\em FLOPS '10: Proceedings of the Tenth International Symposium
  on Functional and Logic Programming}. Springer-Verlag, London, UK.
\newblock To appear.

\bibitem[\protect\citeauthoryear{Danvy and Nielsen}{Danvy and
  Nielsen}{2004}]{danvyTRrefocusing}
{\sc Danvy, O.} {\sc and} {\sc Nielsen, L.~R.} 2004.
\newblock Refocusing in reduction semantics.
\newblock Tech. Rep. RS-04-26, BRICS, DAIMI, Department of Computer Science,
  University of Aarhus, Aarhus, Denmark. November.

\bibitem[\protect\citeauthoryear{Dybvig, Peyton~Jones, and Sabry}{Dybvig
  et~al\mbox{.}}{2007}]{dybvig07monadic}
{\sc Dybvig, R.~K.}, {\sc Peyton~Jones, S.}, {\sc and} {\sc Sabry, A.} 2007.
\newblock A monadic framework for delimited continuations.
\newblock {\em Journal of Functional Programming\/}~{\em 17,\/}~6, 687--730.

\bibitem[\protect\citeauthoryear{Felleisen}{Felleisen}{1988}]{felleisen88promp%
ts}
{\sc Felleisen, M.} 1988.
\newblock The theory and practice of first-class prompts.
\newblock In {\em POPL '88: Proceedings of the 15th ACM SIGPLAN-SIGACT
  Symposium on Principles of Programming Languages}. ACM, New York, NY, USA,
  180--190.

\bibitem[\protect\citeauthoryear{Felleisen, Findler, and Flatt}{Felleisen
  et~al\mbox{.}}{2009}]{findler09redex}
{\sc Felleisen, M.}, {\sc Findler, R.}, {\sc and} {\sc Flatt, M.} 2009.
\newblock {\em Semantics Engineering with {PLT} {Redex}}.
\newblock MIT Press, Cambridge, MA.

\bibitem[\protect\citeauthoryear{Felleisen and Friedman}{Felleisen and
  Friedman}{1986}]{felleisen86secd}
{\sc Felleisen, M.} {\sc and} {\sc Friedman, D.~P.} 1986.
\newblock Control operators, the {SECD}-machine, and the $\lambda$-calculus.
\newblock In {\em Formal Description of Programming Concepts}, {M.~Wirsing},
  Ed. North-Holland, Netherlands, 193--217.

\bibitem[\protect\citeauthoryear{Felleisen and Hieb}{Felleisen and
  Hieb}{1992}]{felleisen92revised}
{\sc Felleisen, M.} {\sc and} {\sc Hieb, R.} 1992.
\newblock A revised report on the syntactic theories of sequential control and
  state.
\newblock {\em Theoretical Computer Science\/}~{\em 103,\/}~2, 235--271.

\bibitem[\protect\citeauthoryear{Friedman, Ghuloum, Siek, and
  Winebarger}{Friedman et~al\mbox{.}}{2007}]{friedman07krivine}
{\sc Friedman, D.~P.}, {\sc Ghuloum, A.}, {\sc Siek, J.~G.}, {\sc and} {\sc
  Winebarger, O.~L.} 2007.
\newblock Improving the lazy {Krivine} machine.
\newblock {\em Higher-Order and Symbolic Computation\/}~{\em 20,\/}~3,
  271--293.

\bibitem[\protect\citeauthoryear{Friedman and Wise}{Friedman and
  Wise}{1976}]{friedman76cons}
{\sc Friedman, D.~P.} {\sc and} {\sc Wise, D.~S.} 1976.
\newblock {CONS} should not evaluate its arguments.
\newblock In {\em Automata, Languages and Programming}, {S.~Michaelson} {and}
  {R.~Milner}, Eds. Edinburgh University Press, Edinburgh, Scotland, 257--284.

\bibitem[\protect\citeauthoryear{Gibbons and Wansbrough}{Gibbons and
  Wansbrough}{1996}]{gibbons96tracing}
{\sc Gibbons, J.} {\sc and} {\sc Wansbrough, K.} 1996.
\newblock Tracing lazy functional languages.
\newblock In {\em Proceedings of Conference on Computing: The Australian Theory
  Symposium}, {M.~E. Houle} {and} {P.~Eades}, Eds. Australian Computer Science
  Communications, Townsville, 11--20.

\bibitem[\protect\citeauthoryear{Henderson and Morris}{Henderson and
  Morris}{1976}]{henderson76lazy}
{\sc Henderson, P.} {\sc and} {\sc Morris, Jr., J.~H.} 1976.
\newblock A lazy evaluator.
\newblock In {\em POPL '76: Proceedings of the 3rd ACM SIGACT-SIGPLAN Symposium
  on Principles of Programming Languages}. ACM, New York, NY, USA, 95--103.

\bibitem[\protect\citeauthoryear{Kameyama, Kiselyov, and Shan}{Kameyama
  et~al\mbox{.}}{2008}]{kameyama08closing}
{\sc Kameyama, Y.}, {\sc Kiselyov, O.}, {\sc and} {\sc Shan, C.} 2008.
\newblock Closing the stage: From staged code to typed closures.
\newblock In {\em PEPM '08: Proceedings of the 2008 ACM SIGPLAN Symposium on
  Partial Evaluation and Semantics-based Program Manipulation}. ACM, New York,
  NY, USA, 147--157.

\bibitem[\protect\citeauthoryear{Kiselyov, Shan, Friedman, and Sabry}{Kiselyov
  et~al\mbox{.}}{2005}]{kiselyov05backtracking}
{\sc Kiselyov, O.}, {\sc Shan, C.}, {\sc Friedman, D.~P.}, {\sc and} {\sc
  Sabry, A.} 2005.
\newblock Backtracking, interleaving, and terminating monad transformers:
  (functional pearl).
\newblock In {\em ICFP '05: Proceedings of the Tenth ACM SIGPLAN International
  Conference on Functional Programming}. ACM, New York, NY, USA, 192--203.

\bibitem[\protect\citeauthoryear{Kiselyov, Shan, and Sabry}{Kiselyov
  et~al\mbox{.}}{2006}]{kiselyov06dynamic}
{\sc Kiselyov, O.}, {\sc Shan, C.}, {\sc and} {\sc Sabry, A.} 2006.
\newblock Delimited dynamic binding.
\newblock In {\em ICFP '06: Proceedings of the eleventh ACM SIGPLAN
  international conference on Functional programming}. ACM, New York, NY, USA,
  26--37.

\bibitem[\protect\citeauthoryear{Krivine}{Krivine}{2007}]{krivine07machine}
{\sc Krivine, J.-L.} 2007.
\newblock A call-by-name lambda-calculus machine.
\newblock {\em Higher-Order and Symbolic Computation\/}~{\em 20,\/}~3,
  199--207.

\bibitem[\protect\citeauthoryear{Launchbury}{Launchbury}{1993}]{launchbury93na%
tural}
{\sc Launchbury, J.} 1993.
\newblock A natural semantics for lazy evaluation.
\newblock In {\em POPL '93: Proceedings of the 20th ACM SIGPLAN-SIGACT
  Symposium on Principles of Programming Languages}. ACM, New York, NY, USA,
  144--154.

\bibitem[\protect\citeauthoryear{Maraist, Odersky, and Wadler}{Maraist
  et~al\mbox{.}}{1998}]{maraist98need}
{\sc Maraist, J.}, {\sc Odersky, M.}, {\sc and} {\sc Wadler, P.} 1998.
\newblock The call-by-need lambda calculus.
\newblock {\em Journal of Functional Programming\/}~{\em 8,\/}~3 (May),
  275--317.

\bibitem[\protect\citeauthoryear{Moggi and Sabry}{Moggi and
  Sabry}{2004}]{sabry04recursion}
{\sc Moggi, E.} {\sc and} {\sc Sabry, A.} 2004.
\newblock An abstract monadic semantics for value recursion.
\newblock {\em Theoretical Informatics and Applications\/}~{\em 38,\/}~4,
  375--400.

\bibitem[\protect\citeauthoryear{Nakata and Hasegawa}{Nakata and
  Hasegawa}{2009}]{nakata10need}
{\sc Nakata, K.} {\sc and} {\sc Hasegawa, M.} 2009.
\newblock Small-step and big-step semantics for call-by-need.
\newblock {\em Journal of Functional Programming\/}~{\em 19,\/}~06, 699--722.

\bibitem[\protect\citeauthoryear{Okasaki, Lee, and Tarditi}{Okasaki
  et~al\mbox{.}}{1994}]{OKASAKI94}
{\sc Okasaki, C.}, {\sc Lee, P.}, {\sc and} {\sc Tarditi, D.} 1994.
\newblock Call-by-need and continuation-passing style.
\newblock {\em Lisp and Symbolic Computation\/}~{\em 7,\/}~1 (Jan.), 57--81.

\bibitem[\protect\citeauthoryear{Peyton~Jones and Salkild}{Peyton~Jones and
  Salkild}{1989}]{peytonjones89spineless}
{\sc Peyton~Jones, S.~L.} {\sc and} {\sc Salkild, J.} 1989.
\newblock The spineless tagless {G}-machine.
\newblock In {\em FPCA '89: Proceedings of the Fourth International Conference
  on Functional Programming Languages and Computer Architecture}. ACM, New
  York, NY, USA, 184--201.

\bibitem[\protect\citeauthoryear{Plotkin}{Plotkin}{1975}]{plotkin75byname}
{\sc Plotkin, G.~D.} 1975.
\newblock Call-by-name, call-by-value and the {$\lambda$}-calculus.
\newblock {\em Theoretical Computer Science\/}~{\em 1,\/}~2 (Dec.), 125--159.

\bibitem[\protect\citeauthoryear{Sestoft}{Sestoft}{1997}]{sestoft97machine}
{\sc Sestoft, P.} 1997.
\newblock Deriving a lazy abstract machine.
\newblock {\em Journal of Functional Programming\/}~{\em 7,\/}~3, 231--264.

\bibitem[\protect\citeauthoryear{Shan}{Shan}{2007}]{shan07static}
{\sc Shan, C.} 2007.
\newblock A static simulation of dynamic delimited control.
\newblock {\em Higher-Order and Symbolic Computation\/}~{\em 20,\/}~4,
  371--401.

\bibitem[\protect\citeauthoryear{Sussman and {Steele Jr.}}{Sussman and {Steele
  Jr.}}{1975}]{sussman98scheme}
{\sc Sussman, G.~J.} {\sc and} {\sc {Steele Jr.}, G.~L.} 1975.
\newblock Scheme: An interpreter for extended lambda calculus.
\newblock AI Memo 349, Artificial Intelligence Laboratory, Massachusetts
  Institute of Technology, Cambridge, Massachusetts. Dec.
\newblock Reprinted in Higher-Order and Symbolic Computation 11(4):405--439,
  1998, with a foreword~\cite{Sussman-Steele:HOSC98-revisited}.

\bibitem[\protect\citeauthoryear{Sussman and {Steele Jr.}}{Sussman and {Steele
  Jr.}}{1998}]{Sussman-Steele:HOSC98-revisited}
{\sc Sussman, G.~J.} {\sc and} {\sc {Steele Jr.}, G.~L.} 1998.
\newblock The first report on {S}cheme revisited.
\newblock {\em Higher-Order and Symbolic Computation\/}~{\em 11,\/}~4,
  399--404.

\bibitem[\protect\citeauthoryear{Wand and Vaillancourt}{Wand and
  Vaillancourt}{2004}]{wand04backtracking}
{\sc Wand, M.} {\sc and} {\sc Vaillancourt, D.} 2004.
\newblock Relating models of backtracking.
\newblock In {\em ICFP '04: Proceedings of the Ninth {ACM SIGPLAN}
  International Conference on Functional Programming}. ACM, New York, NY, USA,
  54--65.

\bibitem[\protect\citeauthoryear{Wang}{Wang}{1990}]{wang90lazy}
{\sc Wang, C.} 1990.
\newblock Obtaining lazy evaluation with continuations in {Scheme}.
\newblock {\em Information Processing Letters\/}~{\em 35,\/}~2, 93--97.

\bibitem[\protect\citeauthoryear{Xi}{Xi}{1997}]{xi97underlambda}
{\sc Xi, H.} 1997.
\newblock Evaluation under lambda abstraction.
\newblock In {\em PLILP '97: Proceedings of the Ninth International Symposium
  on Programming Languages: Implementations, Logics, and Programs}.
  Springer-Verlag, London, UK, 259--273.

\end{thebibliography}
